\documentstyle[11pt,twoside,twocolumn,game,epsf,rotate,epic]{spiel}
\newcounter{saveeqn}                  
\newcommand{\alpheqn}[1]{\refstepcounter{equation}\label{#1}%
\setcounter{saveeqn}{\value{equation}}%
\setcounter{equation}{0}%
\renewcommand{\theequation}
{\mbox{\arabic{saveeqn}\alph{equation}}}}
\newcommand{\reseteqn}{\setcounter{equation}{\value{saveeqn}}%
\renewcommand{\theequation}{\arabic{equation}}}
\newcounter{savefig}
\newcommand{\alphfig}[1]{\refstepcounter{figure}\label{#1}%
\setcounter{savefig}{\value{figure}}%
\setcounter{figure}{0}%
\renewcommand{\thefigure}
{\mbox{\arabic{savefig}\alph{figure}}}}
\newcommand{\resetfig}{\setcounter{figure}{\value{savefig}}%
\renewcommand{\thefigure}{\arabic{figure}}}
\newcommand{\capt}[1]{\newcommand{\groesse}{\normalsize}%
\renewcommand{\normalsize}{\footnotesize}%
\caption[ ]{#1}%
\renewcommand{\normalsize}{\groesse}}

\renewcommand{\bibitem}[1]{\\[0.2cm]\vphantom{#1}}
\renewcommand{\vec}[1]{\mbox{\boldmath $#1$}}
\newcommand{\weg}[1]{#1}                       
\begin{document}
\setcounter{page}{117}
\pagestyle{myheadings}                                    
\title{A Mathematical Model for 
Behavioral Changes by Pair Interactions
and Its Relation to Game Theory}

\author{Dirk Helbing} 
\maketitle
\begin{abstract}
A mathematical mo\-del for be\-ha\-vio\-ral chan\-ges by pair interactions
(i.e. due to direct contact) of individuals is developed. Three kinds of
pair interactions can be distinguished: Imitative processes,
avoidance processes, and compromising
processes. Representative solutions of the model
for two different interacting subpopulations are illustrated by
computational results.
\par
The equations of game theory are shown to result for 
a special case of imitative
processes. Moreover, a stochastic version of game theory is formulated.
It allows the derivation of equations for the most probable or the
expected distribution of behavioral strategies and of (co)variance equations. 
The knowledge of the (co)variances is necessary for the calculation of the
reliability of game theoretical descriptions. 
\par
The use and application of the introduced equations is illustrated by
concrete examples. Especially, computational
results for the selforganization of social conventions by
competition of two equivalent strategies are presented.
\end{abstract}

\section{Introduction}

This paper treats a mathematical model for the change of the fraction
$P(i,t)$ of individuals who show a certain behavior $i$. Models of this
kind are of great interest for a {\em quantitative understanding}
or {\em prognosis} of social developments. For the description of the
competition or cooperation of populations there already exist
{\em game theoretical approaches} (see, for example, {\sc Mueller} (1990),
{\sc Axelrod} (1984), {\sc von Neumann} and {\sc Morgenstern}
(1944), {\sc Luce} and {\sc Raiffa} (1957)). However,
the model devoloped in this paper shows to be more general,
since it includes as special cases 
\begin{itemize}
\item not only the {\em game dynamical equations} ({\sc Hofbauer}
and {\sc Sigmund} (1988)), but also
\item the {\em logistic equation} ({\sc Verhulst} (1845), {\sc Pearl} (1924),
{\sc Helbing} (1992)),
\item the {\em Gravity model} ({\sc Ravenstein} (1876), 
{\sc Zipf} (1946)),
\item the {\sc Lotka-Vol\-ter\-ra} equa\-tions\linebreak 
({\sc Lot\-ka} (1920, 1956), 
{\sc Vol\-ter\-ra} (1931), {\sc Hof\-bauer} (1981),
{\sc Goel} et. al. (1971), {\sc Hallam} (1986), {\sc Good\-win} (1967)), and
\item the quantitative social models of {\sc Weid\-lich} and {\sc Haag} 
({\sc Weid\-lich} \& {\sc Haag} (1983, 1988), {\sc Weid\-lich} (1991)).
\end{itemize}
This model assumes behavioral changes to occur with a certain {\em probability}
per time unit, called the {\em transition rate}. The transition rate is
decomposed into                                                        
\begin{itemize}
\item a rate describing {\em spontaneous} behavioral changes, and
\item a rate describing behavioral changes due to {\em pair interactions}
of individuals.
\end{itemize}
Three different kinds of pair interactions can be distinguished: 
\begin{itemize}
\item First, {\em imitative processes}, 
which describe the tendency to take over
the behavior of another individual.
\item Second, {\em avoidance processes}, causing an individual to change the 
behavior if meeting another individual with the same behavior.
\item Third, {\em compromising processes}, 
which describe the readiness to change the 
behavior to a new one when meeting an individual with another behavior.
\end{itemize}
Representative solutions of the model are illustrated by computer simulations.
By distinguishing several {\em subpopulations} $a$, 
different {\em types} of behavior
can be taken into account.
\par
As one would expect, there is a connection of this model with the 
game dynamical equations. In order to establish this connection, the
transition rates have to be taken in a special way which depends on the
{\em expected success} of the behavioral strategies. The essential effect
is given by imitative processes.
\par
A stochastic formulation of the game dynamic equations shows that the
ordinary game dynamical equations result as equations for the {\em most
probable} behavioral distribution 
or as {\em approximate mean value} equations. 
\par
For the approximate mean values corrections can be calculated. These
corrections depend on the (co)variances $\sigma_{ij}$
of the numbers $n_i$ of individuals who show a certain behavior $i$. 
The calculation of the (co)variances is also useful to determine
the reliability of game theoretical descriptions. 
\par
An example
of two equivalent competing strategies serves as an illustration of the
game dynamical equations and their stochastic version. It allows the
description of the {\em selforganization} of a behavioral convention.

\section{The master equation} \label{stoch}

Suppose, we have a social system with $N$ individuals. These individuals
can be divided into $A$ {\em subpopulations} $a$ consisting of 
$N_a$ individuals, i.e.,
\begin{displaymath}
 \sum_{a=1}^A N_a = N \, .
\end{displaymath}
By subpopulations different social groups (e.g. blue and white 
collars) or different characteristic {\em types} of behavior are
distinguished. 
\par
The $N_a$ individuals of each subpopulation $a$ are distributed
over several {\em states}
\begin{displaymath}
 i \in \{1,\dots,S\} \, ,
\end{displaymath}
which represent the {\em behavior} or the (behavioral) 
{\em strategy} of an individual. If the {\em occupation number} $n_i^a$
denotes the number of individuals of subpopulation $a$ who show the 
behavior $i$, we have the relation
\begin{equation}
 \sum_{i=1}^S n_i^a = N_a \, .
\label{sum}
\end{equation}
Let 
\begin{displaymath}
 \vec{n} := (n_1^1,\dots,n_i^a,\dots,n_S^A)
\end{displaymath}
be the vector consisting of all occupation numbers $n_i^a$. This vector is
called the {\em socioconfiguration}, since it contains all information
about the distribution of the $N$ individuals over the states $i$. 
$P(\vec{n},t)$ shall denote the {\em probability} to find the
socioconfiguration $\vec{n}$ at time $t$. This implies
\begin{displaymath}
 0 \le P(\vec{n},t) \le 1 \quad \mbox{and} \quad
\sum_{\weg{n}} P(\vec{n},t)= 1 \, .
\end{displaymath}
If transitions
from socioconfiguration $\vec{n}$ to $\vec{n}'$
occur with a probability of $P(\vec{n}',t+\Delta t|\vec{n},t)$ 
during a short time interval
$\Delta t$, we have a {\em (relative) transition rate} of
\begin{displaymath}
 w(\vec{n}',\vec{n};t) 
:= \lim_{\Delta t \rightarrow 0}
\frac{P(\vec{n}',t+\Delta t|\vec{n},t)}{\Delta t} \, .
\end{displaymath}
The {\em absolute} transition rate of changes
from $\vec{n}$ to $\vec{n}'$ is the product 
$w(\vec{n}',\vec{n};t)P(\vec{n},t)$ of the probability
$P(\vec{n},t)$ to have the configuration $\vec{n}$
and the {\em relative} transition rate $w(\vec{n}',\vec{n};t)$
if having the configuration $\vec{n}$. Whereas the {\em inflow} into
$\vec{n}$ is given as
the sum over all absolute transition rates of changes from an {\em arbitrary}
configuration $\vec{n}'$ to $\vec{n}$, the {\em outflow} from $\vec{n}$
is given as the sum over all absolute transition rates of changes
from $\vec{n}$ to {\em another} configuration $\vec{n}'$. Since the
temporal change of the probability $P(\vec{n},t)$ is determined
by the inflow into $\vec{n}$ reduced by the outflow from $\vec{n}$, we
find the {\em master equation}\pagebreak
\begin{eqnarray}
 \frac{d}{dt} P(\vec{n},t) &=&
\mbox{inflow into $\vec{n}$ } \nonumber \\
&-& \mbox{outflow from $\vec{n}$} \nonumber \\
&=& \sum_{\weg{n}'} w(\vec{n},\vec{n}';t)P(\vec{n}',t) \nonumber \\
&-& \sum_{\weg{n}'} w(\vec{n}',\vec{n};t)P(\vec{n},t) \qquad
\label{master}
\end{eqnarray}
(see {\sc Haken} (1983)), which is a {\em 
stochastic equation}.
\par
It shall be assumed that two processes contribute to a change of the
socioconfiguration $\vec{n}$:
\begin{itemize}
\item Individuals may change their behavior $i$ spontaneously and
independently of each other to another behavior $i'$ with an
{\em individual} transition rate $\widetilde{w}_a(i',i;t)$. 
These changes correspond to transitions of the socioconfiguration from
$\vec{n}$ to                                                                  
\begin{eqnarray*}
 \vec{n}_{i'i}^a &:=& (n_1^1,\dots,(n_{i'}^a + 1), \nonumber \\
&\times& \dots, (n_i^a -1), \dots,n_S^A)
\end{eqnarray*}
with a {\em configurational} transition rate $w(\vec{n}_{i'i}^a,\vec{n};t)=
n_i^a \widetilde{w}_a(i',i;t)$, which is
proportional to the number $n_i^a$ of individuals who can change the
behavior $i$.
\item An individual of subpopulation $a$ may change the behavior from
$i$ to $i'$ during a pair interaction with 
an individual of a subpopulation
$b$ who changes the behavior from $j$ to $j'$. Let transitions of this kind
occur with a probability $\widetilde{w}_{ab}(i',j';i,j;t)$ per time unit. The
corresponding change of the socioconfiguration from $\vec{n}$ to
\begin{eqnarray*}
 \vec{n}_{i'j'ij}^{ab} 
&:=& (n_1^1,\dots,(n_{i'}^a + 1),\dots, \nonumber \\
&\times& (n_i^a -1),\dots,(n_{j'}^b +1), \nonumber \\
&\times& \dots,(n_j^b-1),\dots,n_S^A)
\end{eqnarray*}
leads to a configurational transition rate
$w(\vec{n}_{i'j'ij}^{ab},\vec{n};t)
= n_i^a n_j^b \widetilde{w}_{ab}(i',j';i,j;t)$,
which is proportional to the number
$n_i^an_j^b$ of possible pair interactions between individuals of
subpopulations $a$ resp. $b$ who show the behavior $i$ resp. $j$.
(Exactly speaking---in order to exclude 
self-interactions---\mbox{$n_i^an_i^a\widetilde{w}_{aa}(i',j';i,i;t)$} 
has to be replaced by
\mbox{$n_i^a(n_i^a-1)\widetilde{w}_{aa}(i',j';i,i;t)$}, if $P(\vec{n},t)$ is
not negligible where $n_i^a \gg 1$ does not hold, and
$\sum_{j'}\widetilde{w}_{aa}(i',j';i,i;t)
\ll \widetilde{w}_a(i',i;t)$ is invalid.)
\end{itemize}
The resulting configurational transition rate $w(\vec{n}',\vec{n};t)$ is
given by
\begin{eqnarray}
\!\!\!\!& &\!\!w(\vec{n}',\vec{n};t) \vphantom{\int} \nonumber \\
\!\!\!\!&:=&\!\!\left\{
\begin{array}{ll}
n_i^a \widetilde{w}_a(i',i;t) & \mbox{if } \vec{n}' = \vec{n}_{i'i}^a \\
n_i^a n_j^b \widetilde{w}_{ab}(i',j';i,j;t) & \mbox{if }
\vec{n}' = \vec{n}_{i'j'ij}^{ab} \\
0 & \mbox{otherwise.}
\end{array}\right. \nonumber \\
\!\!\!\!& & 
\label{rate}
\end{eqnarray}
As a consequence, the explicit form of the master equation (\ref{master})
is
\begin{eqnarray*}
& & \frac{d}{dt}P(\vec{n},t) \nonumber \\
&=& \sum_{a,i,i'}
\Big[ (n_{i'}^a+1)\widetilde{w}_a(i,i';t)P(\vec{n}_{i'i}^a,t) \nonumber \\
& & \qquad - \; n_i^a \widetilde{w}_a(i',i;t)P(\vec{n},t) \Big] \nonumber \\
&+& \frac{1}{2} \sum_{a,i,i'} \sum_{b,j,j'} \Big[
(n_{i'}^a+1)(n_{j'}^b+1)\nonumber \\
& & \qquad \quad \times 
\widetilde{w}_{ab}(i,j;i',j';t)P(\vec{n}_{i'j'ij}^{ab},t) \vphantom{\sum_b}
\nonumber \\
& &\qquad - \; n_i^a n_j^b \widetilde{w}_{ab}(i',j';i,j;t)P(\vec{n},t) \Big] 
\end{eqnarray*}                          
(see {\sc Helbing} (1992a)).

\section{Most probable and expected behavioral distribution}

Because of the great number of possible socioconfigurations $\vec{n}$,
the master equation for the determination of the configurational distribution
$P(\vec{n},t)$ is usually difficult to solve (even with a computer). However,
\begin{itemize}
\item in cases of the description of single or rare social 
processes the {\em most
probable} behavioral distribution
\begin{displaymath}
 P_a(i,t) := \frac{\widehat{n}_i^a(t)}{N_a}
\end{displaymath}
is the quantity of interest, whereas
\pagebreak
\item in cases of frequently occuring social processes the interesting
quantity is the {\em expected} behavioral distribution
\begin{displaymath}
 P_a(i,t) := \frac{\langle n_i^a \rangle_t}{N_a} \, .
\end{displaymath}
\end{itemize}
Equations for the most probable occupation numbers 
$\widehat{n}_i^a(t)$ can be deduced from a {\sc Langevin} equation
for the development of the socioconfiguration $\vec{n}(t)$. 
For the mean values $\langle n_i^a
\rangle_t$ of the occupation numbers
$n_i^a$ only {\em approximate} closed equations can be derived.
A measure for the reliability (or representativity) of $\widehat{n}_i^a(t)$
and $\langle n_i^a \rangle_t$ with respect to the possible temporal
developments of 
$n_i^a(t)$ are the (co)variances $\sigma_{ij}^{ab}(t)$ of 
$n_i^a(t)$.

\subsection{Mean value and (co)variance equations}

The {\em mean value}
of a function $f(\vec{n},t)$ is defined by
\begin{displaymath}
 \langle f(\vec{n},t) \rangle_t \equiv
 \langle f(\vec{n},t) \rangle := \sum_{\weg{n}} f(\vec{n},t) P(\vec{n},t) \, .
\end{displaymath}
It can be shown that the mean values of the occupation 
numbers $f(\vec{n},t) = n_i^a$
are determined by the equations
\begin{equation}
 \frac{d\langle n_i^a\rangle}{dt} = \langle m_i^a(\vec{n},t) \rangle
\label{mean}
\end{equation}
with the {\em drift coefficients}
\begin{eqnarray}
 m_i^a(\vec{n},t) &:=& \sum_{\weg{n}'} (n'{}_i^a - n_i^a)w(\vec{n}',\vec{n};t)
\nonumber \\
&=& \sum_{i'} \Big[ \overline{w}^a(i,i';t)n_{i'}^a \nonumber \\
& & - \; \overline{w}^a(i',i;t) n_i^a \Big]
\label{mean1}
\end{eqnarray}
and the {\em effective transition rates}
\begin{eqnarray}
& & \overline{w}^a(i',i;t) \; := \; \widetilde{w}_a(i',i;t) 
\vphantom{\int} \nonumber \\
&+& \sum_b \sum_{j'}\sum_{j}
\widetilde{w}_{ab}(i',j';i,j;t)n_{j}^b \qquad
\label{mean2}
\end{eqnarray}
(see {\sc Helbing} (1992a)). Obviously, the contributions 
$\widetilde{w}_{ab}(i',j';i,j;t)n_{j}^b$ 
due to pair interactions are proportional
to the number $n_{j}^b$ of possible interaction partners.

\subsubsection{Approximate mean value and (co)variance equations}
\begin{figure}[htbp]
\epsfysize=7.2cm 
\centerline{\rotate[r]{\hbox{\epsffile[57 40 570 756]
{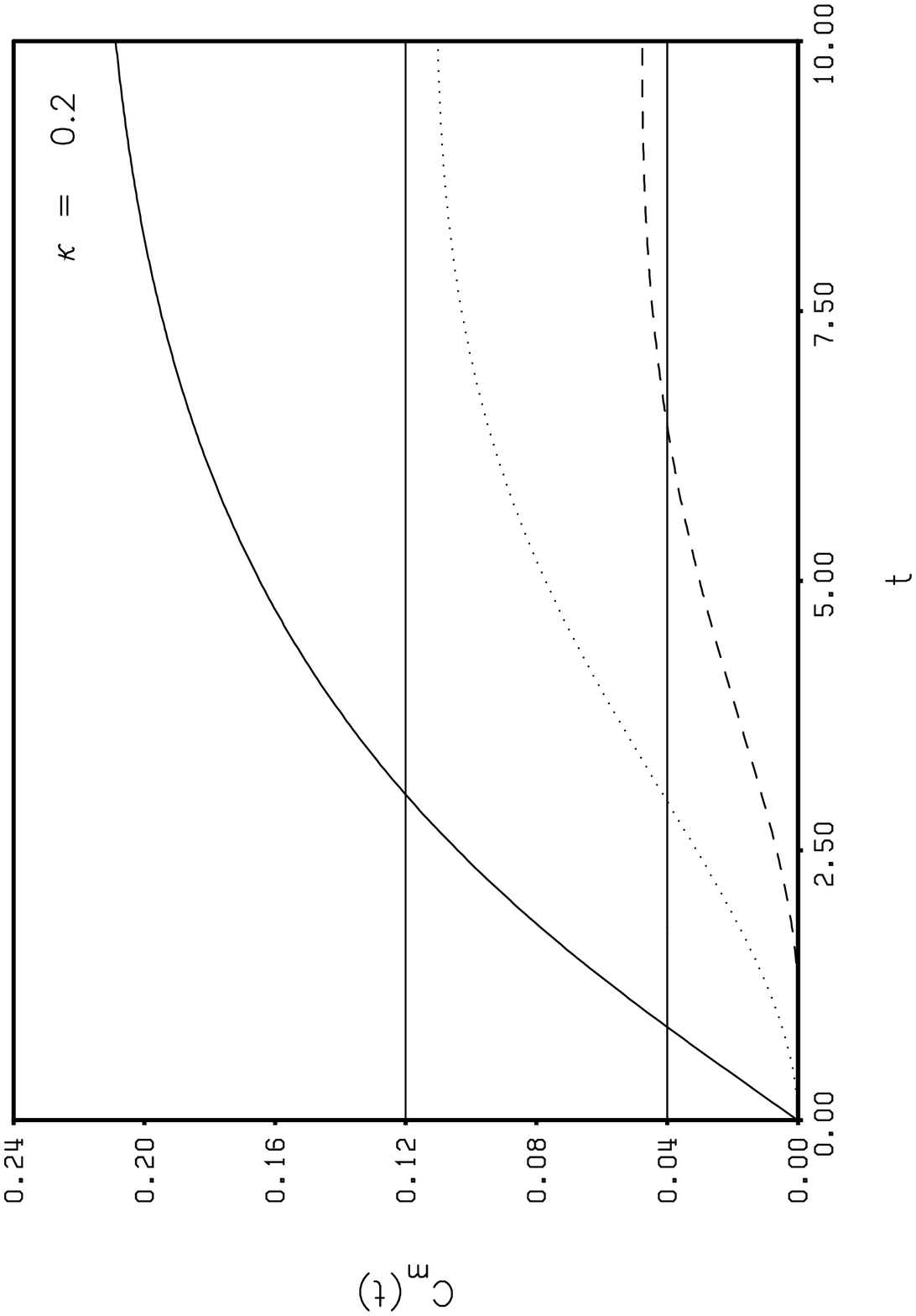}}}}
\capt{The relative central moments are a criterium for the validity
of the approximate resp. the corrected mean value and (co)variance
equations: If $|C_2|$ (---) exceeds the value 0.04, the approximate
equations fail, whereas the corrected equations fail, if $|C_3|$ (-- --)
or $|C_4|$ ($\cdots$) exceed the value 0.04.\label{valid2}}
\end{figure}

\vfill
Equations (\ref{mean}) are no closed equations, since they depend on the
mean values $\langle n_i^a n_j^b \rangle$, which are not determined
by (\ref{mean}). However, if the configurational 
distribution $P(\vec{n},t)$ has only
small (co)variances 
\begin{eqnarray}
 \sigma_{ij}^{ab} &:=& \Big\langle (n_i^a - \langle n_i^a \rangle)
 (n_j^b - \langle n_j^b \rangle) \Big\rangle \nonumber \\
&=& \langle n_i^a n_j^b \rangle - \langle n_i^a \rangle \langle n_j^b \rangle
 \, , \label{Covar}
\end{eqnarray}
we find in first order {\sc Taylor} approximation
the {\em approximate mean value equations}
\begin{eqnarray}
 \frac{\partial \langle n_i^a \rangle}{\partial t}
&\approx& \Bigg\langle m_i^a(\langle \vec{n} \rangle,t) \nonumber \\
&+& \sum_{b,j} (n_j^b - \langle n_j^b \rangle )
\frac{\partial m_i^a(\langle \vec{n} \rangle,t)}{\partial \langle n_j^b\rangle}
\Bigg\rangle \nonumber \\
&=& m_i^a(\langle\vec{n}\rangle,t) \, .
\label{Mean}
\end{eqnarray}
In many cases, the initial configuration $\vec{n}_0$ at time $t_0$ is 
known by a measurement, i.e., the initial distribution is
\begin{displaymath}
 P(\vec{n},t_0) = \delta_{\weg{n}\weg{n}_0} \, ,
\end{displaymath}
where the {\sc Kronecker} function $\delta_{xy}$ is defined by
\begin{displaymath}
 \delta_{xy} := \left\{
\begin{array}{ll}
1 & \mbox{if } x=y \\
0 & \mbox{if } x\ne y \,.
\end{array}\right.
\end{displaymath}
As a consequence, the (co)variances $\sigma_{ij}^{ab}$ vanish at time 
$t_0$ and remain small during a certain time interval. For the temporal
development of $\sigma_{ij}^{ab}$, the equations 
\begin{eqnarray}
 \frac{d\sigma_{ij}^{ab}}{dt} &=& \Big\langle m_{ij}^{ab}(\vec{n},t)
\Big\rangle \nonumber \\
&+& \Big\langle (n_i^a - \langle n_i^a \rangle )m_j^b(\vec{n},t)
\Big\rangle \nonumber \\
&+& \Big\langle (n_j^b - \langle n_j^b \rangle) m_i^a(\vec{n},t)
\Big\rangle
\label{cov}
\end{eqnarray}
can be found. Here,
\begin{eqnarray*}
& & m_{ij}^{ab}(\vec{n},t) \vphantom{\int} \nonumber \\
&:=& \sum_{\weg{n}'} (n'{}_i^a - n_i^a) 
 (n'{}_j^b - n_j^b) w(\vec{n}',\vec{n};t) 
\end{eqnarray*}
\begin{eqnarray}
&=& \delta_{ab} \bigg( \delta_{ii'}
\sum_{j}\Big[ n_{j}^a \overline{w}^a(i,j;t) \nonumber \\
& & \qquad \quad + \; n_{i}^a \overline{w}^a(j,i;t) \Big] \nonumber \\
& & \qquad - \; \Big[n_{i'}^a \overline{w}^a(i,i';t) \nonumber \\
& & \qquad \quad + \; n_{i}^a \overline{w}^a(i', i;t) \Big]\bigg) \nonumber \\
&+& \sum_{j'}\sum_{j} \Big[ 
n_{j}^a n_{j'}^b \widetilde{w}_{ab}
(i,i';j,j';t) \nonumber \\
& &\qquad + \; n_{i}^a n_{i'}^b \widetilde{w}_{ab}
(j,j';i,i';t) \Big] \vphantom{\sum_j} \nonumber \\
&-& \sum_{j'}\sum_{j} \Big[
n_{i}^a n_{j'}^b  \widetilde{w}_{ab}
(j,i';i,j';t) \nonumber \\
& &\qquad + \; n_{j}^an_{i'}^b \widetilde{w}_{ab}
(i,j';j,i';t) \Big] \, . \qquad
\label{cov1}
\end{eqnarray}
are {\em diffusion coefficients} (see {\sc Helbing} (1992a)).
Equations (\ref{cov}) are, again, no closed equations. However, a first order
{\sc Taylor} approximation of the coefficients $m_{..}^{..}(\vec{n},t)$
leads to the equations
\begin{eqnarray}
 \frac{\partial \sigma_{ij}^{ab}}{\partial t} 
&\approx& m_{ij}^{ab}(\langle n \rangle,t)
\nonumber \\
&+& \sum_{c,k} \Bigg( \sigma_{ik}^{ac} \frac{\partial m_j^b(\langle \vec{n}
\rangle,t)}{\partial \langle n_k^c \rangle} \nonumber \\
& & \quad + \; \sigma_{jk}^{bc} \frac{\partial m_i^a(\langle \vec{n}\rangle,t)}
{\partial \langle n_k^c \rangle} \Bigg) \qquad
\label{Cov}
\end{eqnarray}
(see {\sc Helbing} (1992a)), which are solvable together with (\ref{Mean}).
The {\em approximate (co)variance equations} 
(\ref{Cov}) allow the determination of
the time interval during which the approximate mean value equations
(\ref{Mean}) are valid (see figures \ref{valid1} and \ref{valid2}).
They are also useful for the calculation of the
reliability (or representativity) of descriptions made by (\ref{Mean}).
    
\subsubsection{Corrected mean value and (co)variance equations}

Equations (\ref{Mean}) and (\ref{Cov}) are only valid for the case 
\begin{equation}
 \Big| \sigma_{ij}^{ab} \Big| \ll \langle n_i^a \rangle \langle n_j^b \rangle 
\, , \label{sm}
\end{equation}
where the absolute values of the (co)va\-ri\-an\-ces $\sigma_{ij}^{ab}$
are small, i.e., where the configurational distribution $P(\vec{n},t)$ is
sharply peaked. For increasing (co)variances, a better approximation
of (\ref{mean}), (\ref{cov}) should be taken. A second order {\sc Taylor}
approximation results in the {\em corrected mean value equations}
\begin{eqnarray}
 \frac{\partial \langle n_i^a \rangle}{\partial t}
&\approx& m_i^a(\langle \vec{n} \rangle,t) \nonumber \\
&+& \frac{1}{2}
\sum_{b,j}\sum_{c,k} \sigma_{jk}^{bc} \frac{\partial^2 m_i^a(\langle
\vec{n} \rangle,t)}{\partial \langle n_j^b\rangle\partial \langle
n_k^c \rangle} \qquad 
\label{corrmean}
\end{eqnarray}
and the {\em corrected (co)variance equations}
\begin{eqnarray}
 \frac{d\sigma_{ij}^{ab}}{dt} &\approx& m_{ij}^{ab}(\langle n \rangle,t)
\nonumber \\
&+& \frac{1}{2} \sum_{c,k} \sum_{d,l} \sigma_{kl}^{cd} \frac{
\partial^2 m_{ij}^{ab}(\langle\vec{n}\rangle,t)}{\partial \langle n_k^c
\rangle \partial \langle n_l^d \rangle} \qquad \nonumber \\
&+& \sum_{c,k} \Bigg( \sigma_{ik}^{ac} \frac{\partial m_j^b(\langle \vec{n}
\rangle,t)}{\partial \langle n_k^c \rangle} \nonumber \\
& & \qquad + \; \sigma_{jk}^{bc} 
\frac{\partial m_i^a(\langle \vec{n}\rangle,t)}
{\partial \langle n_k^c \rangle} \Bigg) \, . 
\label{corrcov}
\end{eqnarray}
Note, that the corrected mean value equations explicitly depend on the
(co)variances $\sigma_{ij}^{ab}$, i.e.,
on the {\em fluctuations} due to the stochasticity of
the processes described! They cannot be solved without
solving the (co)variance equations. However, the calculation of the
(co)variances is {\em always} recommendable, since they are a measure for the
reliability (or representativity) of the mean value equations.
\par
\begin{figure}[htbp]
\epsfysize=7.2cm 
\centerline{\rotate[r]{\hbox{\epsffile[57 40 570 756]
{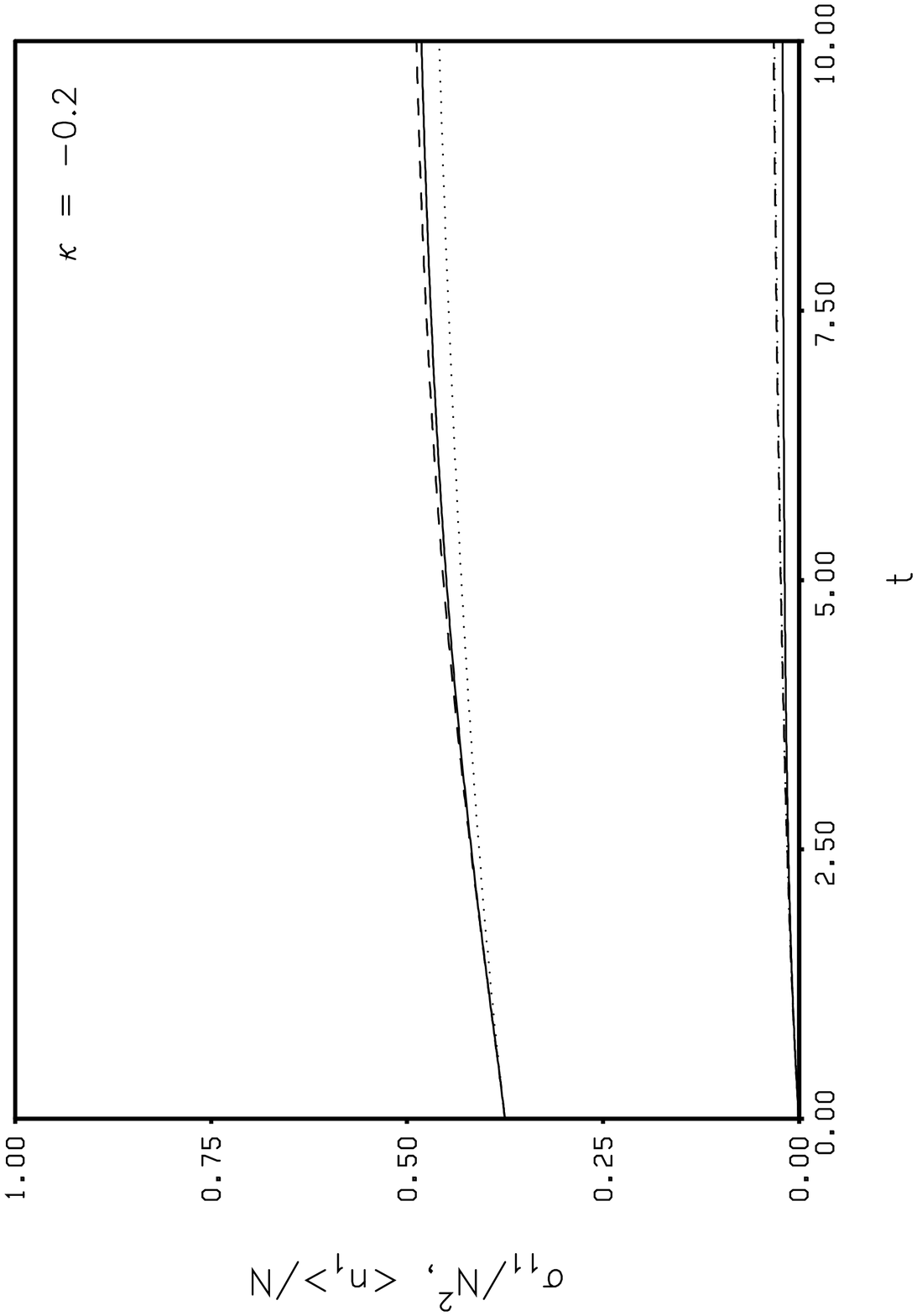}}}}
\capt{Exact (---), approximate ($\cdots$) and corrected (-- --) 
mean values (upper curves) and variances (lower curves)
for a {\em small} configurational distribution $P(\vec{n},t)$:
Both, approximate and corrected equations are applicable.
\label{below2}}
\end{figure}
\begin{figure}[htbp]
\epsfysize=7.2cm 
\centerline{\rotate[r]{\hbox{\epsffile[57 40 570 756]
{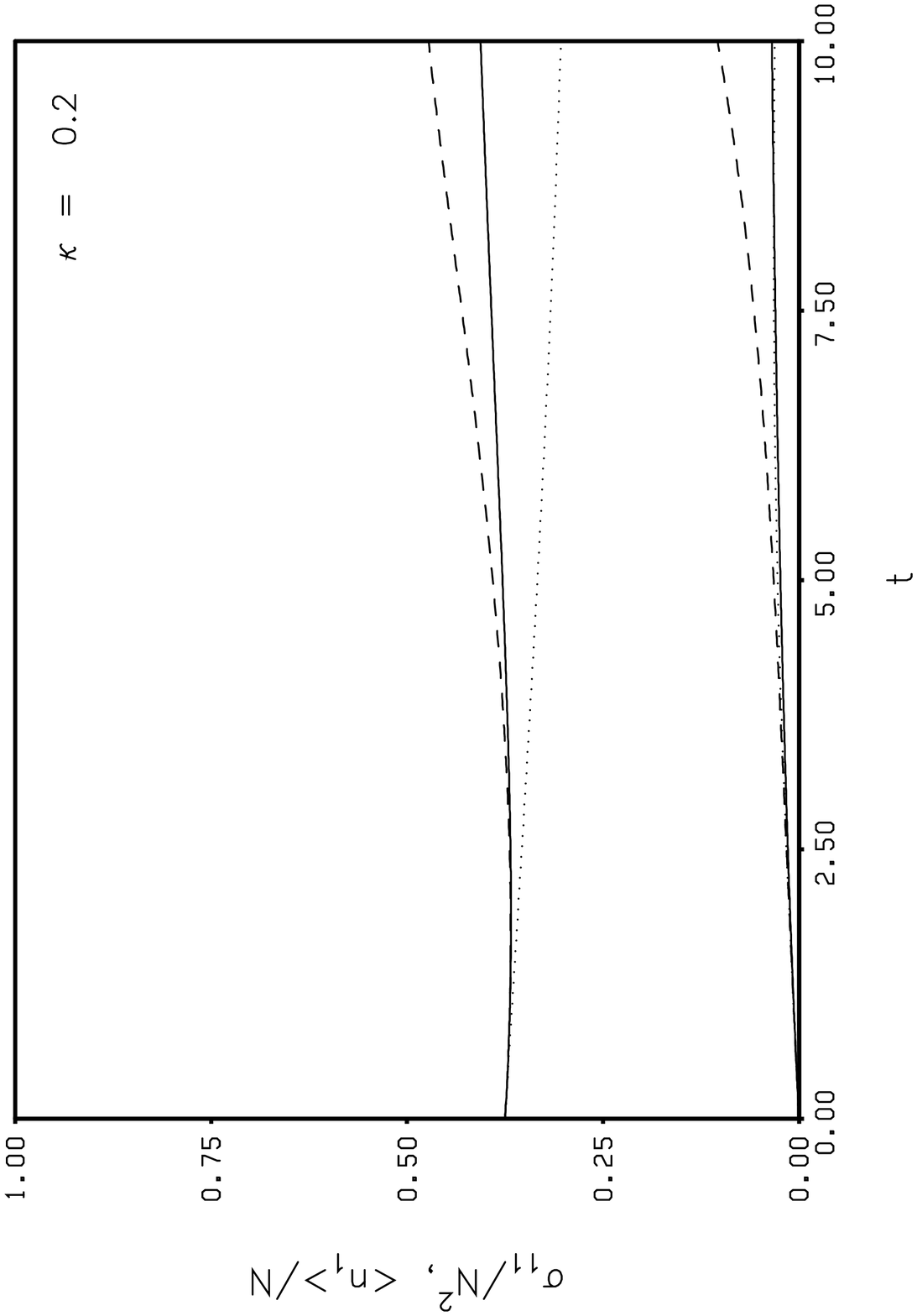}}}}
\capt{As figure \ref{below2}, but for a {\em multimodal} configurational 
distribution: Not only the approximate but also
the corrected equations fail after a certain time interval. 
However, whereas the approximate mean value and variance become unreliable
already for $t > 1$, the corrected mean value and variance remain valid 
until $t>3$.\label{valid1}}
\end{figure}

A comparison of exact, approximate and corrected mean value and variance
equations is 
given in figures \ref{below2} to \ref{valid}. A citerium for the validity of 
(\ref{Mean}) and (\ref{Cov}) resp. (\ref{corrmean}) and (\ref{corrcov})
are the {\em relative central moments}
\begin{eqnarray*}
\!\!\!& &\!\! C_m(i_1,\dots,i_m;t) \vphantom{\int} \nonumber \\
\!\!\!&:=&\!\! \frac{\Big\langle (n_{i_1}^{a_1} - \langle
n_{i_1}^{a_1} \rangle ) \cdot \dots \cdot (n_{i_m}^{a_m} - \langle
n_{i_m}^{a_m} \rangle ) \Big\rangle }{\langle n_{i_1}^{a_1} \rangle
\cdot  \dots  \cdot \langle n_{i_m}^{a_m} \rangle} \, .
\end{eqnarray*}
Whereas the approximate 
equations (\ref{Mean}), (\ref{Cov}) already fail, if
\begin{equation}
 |C_m | \le 0.04
\label{Bed}
\end{equation}
is violated for $m=2$
(compare to (\ref{sm}), (\ref{Covar})), the corrected equations
(\ref{corrmean}), (\ref{corrcov}) only presuppose (\ref{Bed})
for $2 < m \le l$ with a certain value $l$ 
(see {\sc Helbing} (1992a)). However, even
the corrected equations (\ref{corrmean}), (\ref{corrcov}) become useless,
if the probability distribution $P(\vec{n},t)$ becomes multimodal.
\par
Figures \ref{below2} to \ref{valid} show computational results
corresponding to the example of section \ref{beisp}.
{\em Exact} mean values $\langle n_1 \rangle$
and variances $\sigma_{11}$ 
are represented by solid lines, whereas approximate
results according to (\ref{Mean}), (\ref{Cov})
are represented by dotted
lines, and corrected results according to (\ref{corrmean}), (\ref{corrcov})
by broken lines.
As expected, the corrected mean value equations yield better
results than the approximate mean value equations. 
%
%
\vfill
%
\vfill
\begin{figure}[htbp]
\epsfysize=7.2cm 
\centerline{\rotate[r]{\hbox{\epsffile[57 40 570 756]
{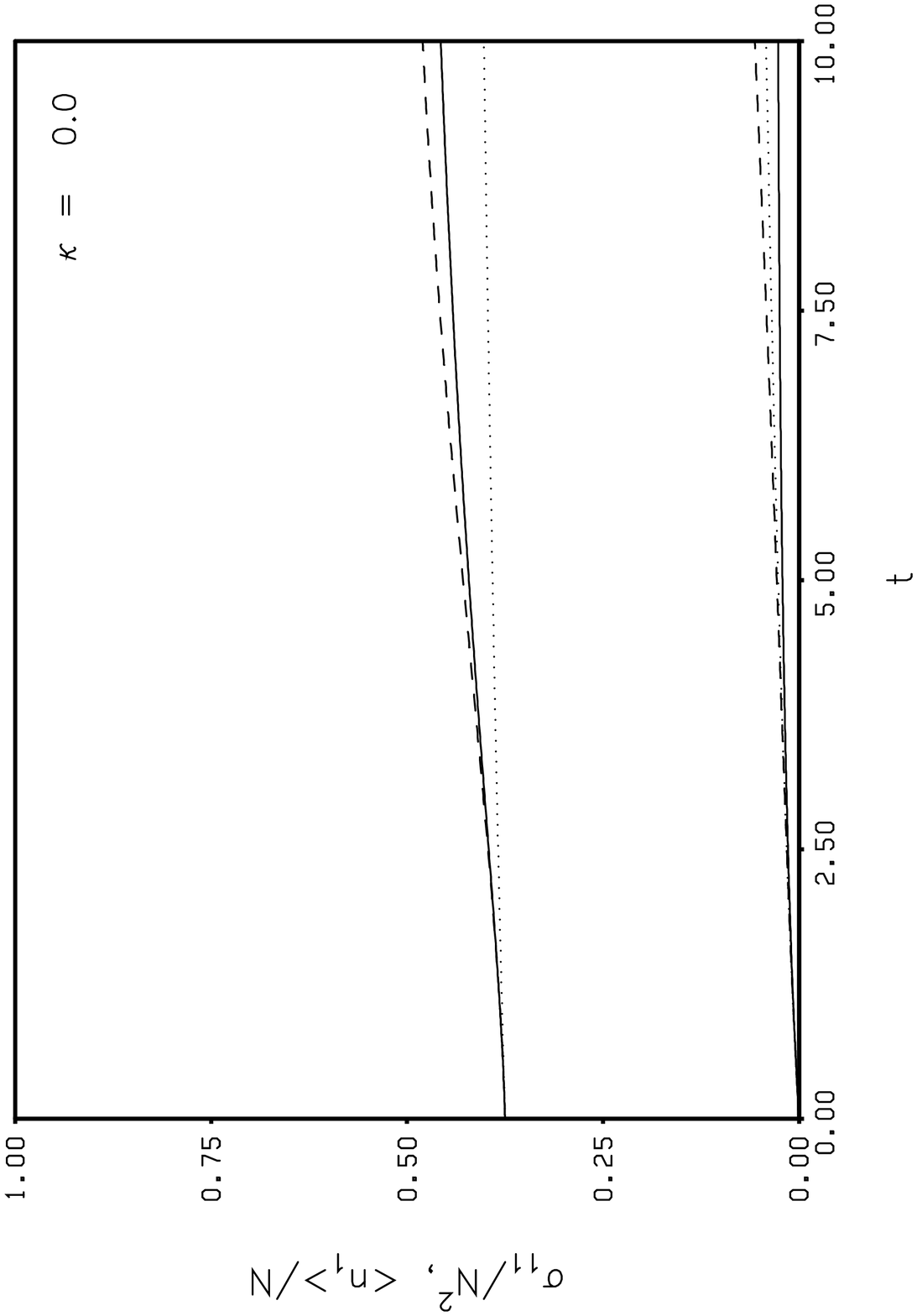}}}}
\capt{As figure \ref{below2}, but for 
a {\em broad} configurational distribution:
The corrected equations still yield
useful results, whereas the approximate equations already fail,
since the variances are not negligible.
\label{at2}}
\end{figure}\alphfig{valid}
\pagebreak
%
%
%
\resetfig

\subsection{Equations for the most probable behavioral distribution}
\label{Most}          
The master equation (\ref{master}) can be reformulated in terms of a
{\sc Langevin} equation (see {\sc Helbing} (1992)):
\begin{equation}
 \frac{d}{dt} n_i^a(t) \stackrel{N \gg 1}{=} m_i^a(\vec{n},t) +
 \mbox{\em fluctuations} \, .
\label{Lang}
\end{equation}
The {\sc Langevin} equation (\ref{Lang}) describes the behavior of the 
socioconfiguration $\vec{n}(t)$ in dependence of 
process immanent fluctuations
(that are determined by the diffusion coefficients $m_{ij}^{ab}$). As a 
consequence,
\begin{equation}
\frac{d}{dt} \widehat{n}_i^a(t) \stackrel{N\gg 1}{=} 
m_i^a(\widehat{\vec{n}},t)
\label{mostprob}
\end{equation}
are the equations for the most probable occupation numbers
$\widehat{n}_i^a(t)$. The equations (\ref{mostprob}) look exactly like the
approximate mean value equations (\ref{Mean}). Therefore,
if $N\gg 1$, the {\em approximate}
mean value equations (\ref{Mean}) have an interpretation even for
great variances, since they also describe the most probable behavioral
distribution. 

\section{Kinds of pair interactions}

The pair interactions
\begin{displaymath}
 i',j' \longleftarrow i,j
\end{displaymath}
of two individuals of subpopulations $a$ resp. $b$ who change their
behavior from $i$ resp. $j$ to $i'$ resp. $j'$ can be completely classified
according to the following scheme:
\begin{displaymath}
\left.
\begin{array}{rcl}
i,i &\longleftarrow & i,i \\
i,j &\longleftarrow & i,j
\end{array} \right\} (0)
\end{displaymath}
\begin{displaymath}
\left.
\begin{array}{rcll}
i,i &\longleftarrow & i,j & (i\ne j)\\
j,j &\longleftarrow & i,j & (i\ne j)
\end{array} \right\} (1)
\end{displaymath}
\begin{displaymath}
\left.
\begin{array}{rcll}
i,j' &\longleftarrow & i,i & (j'\ne i)\\
i',j &\longleftarrow & j,j & (i'\ne j)\\
i',j'&\longleftarrow & i,i & (i'\ne i,j'\ne i)
\end{array} \right\} (2)
\end{displaymath}
\begin{displaymath}
\left.
\begin{array}{rcll}
i,j' &\longleftarrow & i,j & (i\ne j, j'\ne j,j'\ne i)\\
i',j &\longleftarrow & i,j & (i\ne j, i'\ne i, i'\ne j)\\
i',j'&\longleftarrow & i,j & (i\ne j, i'\ne i, j'\ne j,\\
     &               &     & \quad i'\ne j, j'\ne i)
\end{array} \right\} (3)
\end{displaymath}
\begin{displaymath}
\left.
\begin{array}{rcll}
j,i &\longleftarrow & i,j & (i\ne j)\\
i',i &\longleftarrow & i,j & (i\ne j, i'\ne i, i'\ne j)\\
j,j'&\longleftarrow & i,j & (i\ne j, j'\ne j, j'\ne i)
\end{array} \right\} (4)
\end{displaymath}
Obviously, the interpretation of the above {\em kinds}
$k \in \{0,1,\dots,4\}$ of pair interactions is the
following:
\begin{itemize}
\item[(0)] During interactions of kind (0) both individuals do not change
their behavior. These interactions can be omitted in the following, 
since they have no contribution to the change of $P(\vec{n},t)$ or 
$n_i^a(t)$.
\item[(1)] The interactions (1) describe {\em imitative processes}
(processes of persuasion), i.e., the tendency to take over the behavior of
another individual. 
\item[(2)] The interactions (2) describe {\em avoidance processes}, where an
individual changes the behavior when meeting another individual showing the
same behavior. Processes of this kind are known as aversive behavior, defiant
behavior or snob effect.
\item[(3)] The interactions (3) represent some kind of {\em com\-pro\-mi\-sing
pro\-cesses}, where an individual changes the behavior to a new one
(the ``compromise'') when meeting an individual with another behavior.
Such processes are found, if a certain behavior cannot be maintained when
confronted with another behavior.
\item[(4)] The interactions (4) describe imitative processes, in which an 
individual changes the behavior despite of the fact, that he or she 
convinces the interaction partner of his resp. her behavior. Processes of this
kind are very improbable and shall be excluded in the following discussion.
\end{itemize}
For the transition rates corresponding to these kinds of
interaction processes the following plausible form shall be assumed
(see {\sc Helbing} (1992)):
\begin{eqnarray}
\!\!\!\!& &\!\!\!\! \widetilde{w}_{ab}(i',j';i,j;t) 
\; := \; \widetilde{\nu}_{ab}(t) \vphantom{\int} \nonumber \\
\!\!\!\!&\times&\!\!\!\! \left\{
\begin{array}{ll}
p_{ab}^1(i'|i;t)
  & \mbox{if } i'=j \mbox{ and } j' = j\\
p_{ba}^1(j'|j;t)
  & \mbox{if } j'=i \mbox{ and } i' = i\\
0 & \mbox{if } i'=j \mbox{ and } j' \ne j\\
0 & \mbox{if } j'=i \mbox{ and } i' \ne i\\
p_{ab}^k(i'|i;t) & \\
\times p_{ba}^k(j'|j;t) & \mbox{otherwise ($k \in \{2,3\}$).}
\end{array}\right. \nonumber \\
\!\!\!\!& &
\label{TOTRATE}
\end{eqnarray}
Here,
\begin{displaymath}
 \nu_{ab}(t) := N_b \widetilde{\nu}_{ab}(t)
\end{displaymath}
is the {\em contact rate} between an 
individual of subpopulation $a$ with individuals
of subpopulation $b$. $p_{ab}^k(j|i;t)$ is the probability of an individual of
subpopulation $a$ to change the 
behavior from $i$ to $j$ during a pair interaction
of kind $k$ with an individual of subpopulation $b$, i.e.,
\begin{displaymath}
 \sum_j p_{ab}^k (j|i;t) = 1 \, .
\end{displaymath}
Let us assume
\begin{displaymath}
 p_{ab}^k(j|i;t) := f_{ab}^k(t) R_a(j,i;t) \, ,
\end{displaymath}
where $f_{ab}^k(t)$ is a measure for the {\em frequency}
of pair interactions of kind
$k$ between individuals of subpopulation $a$ and $b$, and $R_a(j,i;t)$
is a measure for the {\em readiness} 
of individuals belonging to subpopulation $a$ 
to change the behavior from $i$ to $j$ during a pair interaction. Inserting
the rate (\ref{TOTRATE})
of pair interactions into (\ref{mean2}) and using the conventions
\begin{eqnarray}
 w_a(i',i;t) &:=& \widetilde{w}_a(i',i;t) \, , \nonumber \\
 w_{ab}(i',j';i,j;t) &:=& N_b \widetilde{w}_{ab}(i',j';i,j;t) \, , \nonumber \\
 \nu_{ab}^k(t) &:=& \nu_{ab}(t) f_{ab}^k(t)  \, , \nonumber \\
 P_a(i,t) &:=& \frac{\langle n_i^a \rangle}{N_a} \, , 
\label{ab}
\end{eqnarray}
we arrive at the approximate mean value equations
\begin{eqnarray}
 \frac{d}{dt}P_a(i,t) &=& \sum_{i'} \Big[ w^a(i,i';t)P_a(i',t) \nonumber \\
& &\quad - w^a(i',i;t)P_a(i,t) \Big] \qquad
\label{eff1}
\end{eqnarray}
(see (\ref{Mean}), (\ref{mean1})) with the {\em mean transition rates}
\begin{eqnarray}
 w^a(i,i';t) &:=& w_a(i,i';t) + R_a(i,i';t) \vphantom{\sum_b} \nonumber \\
&\times& \!\!\!\sum_b \Big[ 
\Big( \nu_{ab}^1(t) - \nu_{ab}^3(t) \Big) P_b(i,t) \nonumber \\
& & \!\!\!\! \quad + \Big( \nu_{ab}^2(t) 
- \nu_{ab}^3(t) \Big)P_b(i',t) \vphantom{\sum_b} \nonumber \\
& & \!\!\!\! \quad + \; \nu_{ab}^3(t) \Big] 
\label{eff2}
\end{eqnarray}
(if $N_a \gg 1$; see {\sc Helbing} (1992, 1992b)).
The mean transi\-tion rates in\-clude con\-tri\-bu\-tions 
of spon\-ta\-neous behavioral
changes, and of behavioral changes due to pair interactions (i.e., of 
imitative, avoidance and compromising processes). (\ref{eff1}), (\ref{eff2})
are {\sc Boltzmann}-{\em like equations}
(see {\sc Boltzmann} (1964), {\sc Helbing} (1992a)).
\par
Due to (\ref{sum}), (\ref{ab}), and 
$0 \le n_i^a \le N_a$ we have the relations
\begin{displaymath}
 \sum_i P_a(i,t) = 1 \quad \mbox{and} \quad 0 \le P_a(i,t) \le 1 \, .
\end{displaymath}
Therefore, $P_a(i,t)$ can be interpreted as 
the {\em fraction} of individuals within
subpopulation $a$ who show the behavior $i$. With respect to the {\em total}
population, the fraction $P(i,t)$ of individuals with behavior $i$ is given
by                                             
\begin{eqnarray*}
 P(i,t) &=& \frac{\langle n_i \rangle}{N} \; := \; \frac{\displaystyle \sum_a 
\langle n_i^a \rangle}{N} \nonumber \\
&=& \sum_a \frac{N_a}{N} \frac{\langle n_i^a \rangle}
{N_a} \; = \; \sum_a \frac{N_a}{N} P_a(i,t) \, .
\end{eqnarray*}

\subsection{Computer simulations}

For an illustration of the {\sc Boltzmann}-like equations
(\ref{eff1}), (\ref{eff2}) we shall assume to have two subpopulations
($A=2$), and three different behaviors ($S=3$). With
\begin{equation}
 R_a(i',i;t) := \frac{\mbox{e}^{U_a(i',t) - U_a(i,t)}}{D_a(i',i;t)} \, ,
\label{readiness}
\end{equation}
(see {\sc Weidlich} and {\sc Haag} (1988), {\sc Helbing} (1992))
the readiness $R_a(i',i;t)$ for an individual of subpopulation $a$ to change
the behavior from $i$ to $i'$ will be the greater, the greater the difference
of the {\em utilities} $U_a(.,t)$ 
of behaviors $i'$ and $i$ is, and the smaller the
{\em incompatibility (``distance'')}
\begin{displaymath}
D_a(i',i;t) = D_a(i,i';t) > 0
\end{displaymath}
between the behaviors $i$ and $i'$ is.
\par
\begin{figure}[htbp]
\epsfysize=6.9cm 
\centerline{\rotate[r]{\hbox{\epsffile[57 40 555 756]
{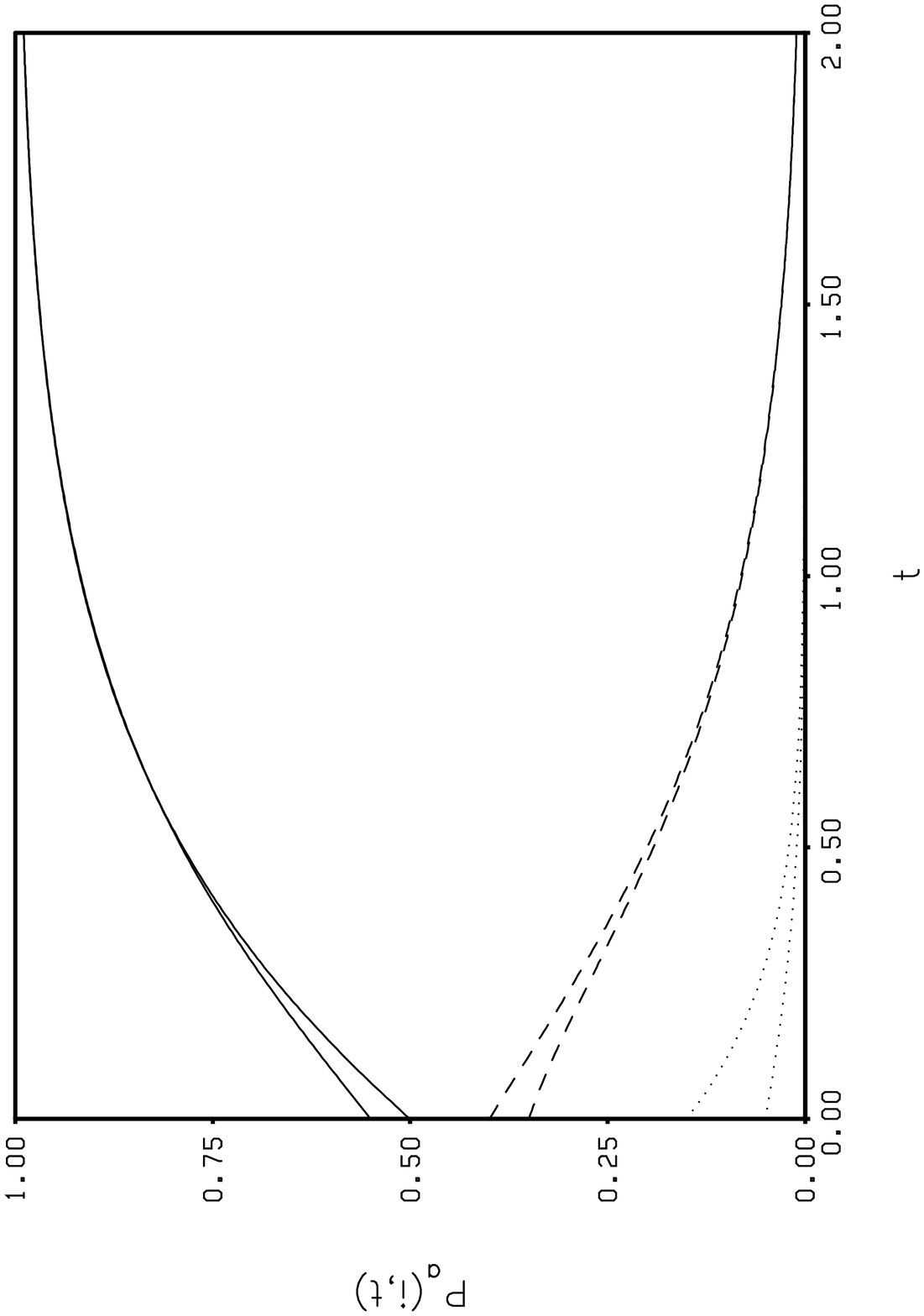}}}}
\capt{Effect of imitative processes for two subpopulations prefering
the same behavior ($c=0.5$):
Only the fraction of the prefered behavior (---) is increasing.
The other behaviors vanish in the course of time.
\label{imit1}}
\end{figure}
\begin{figure}[htbp]
\epsfysize=6.9cm 
\centerline{\rotate[r]{\hbox{\epsffile[57 40 555 756]
{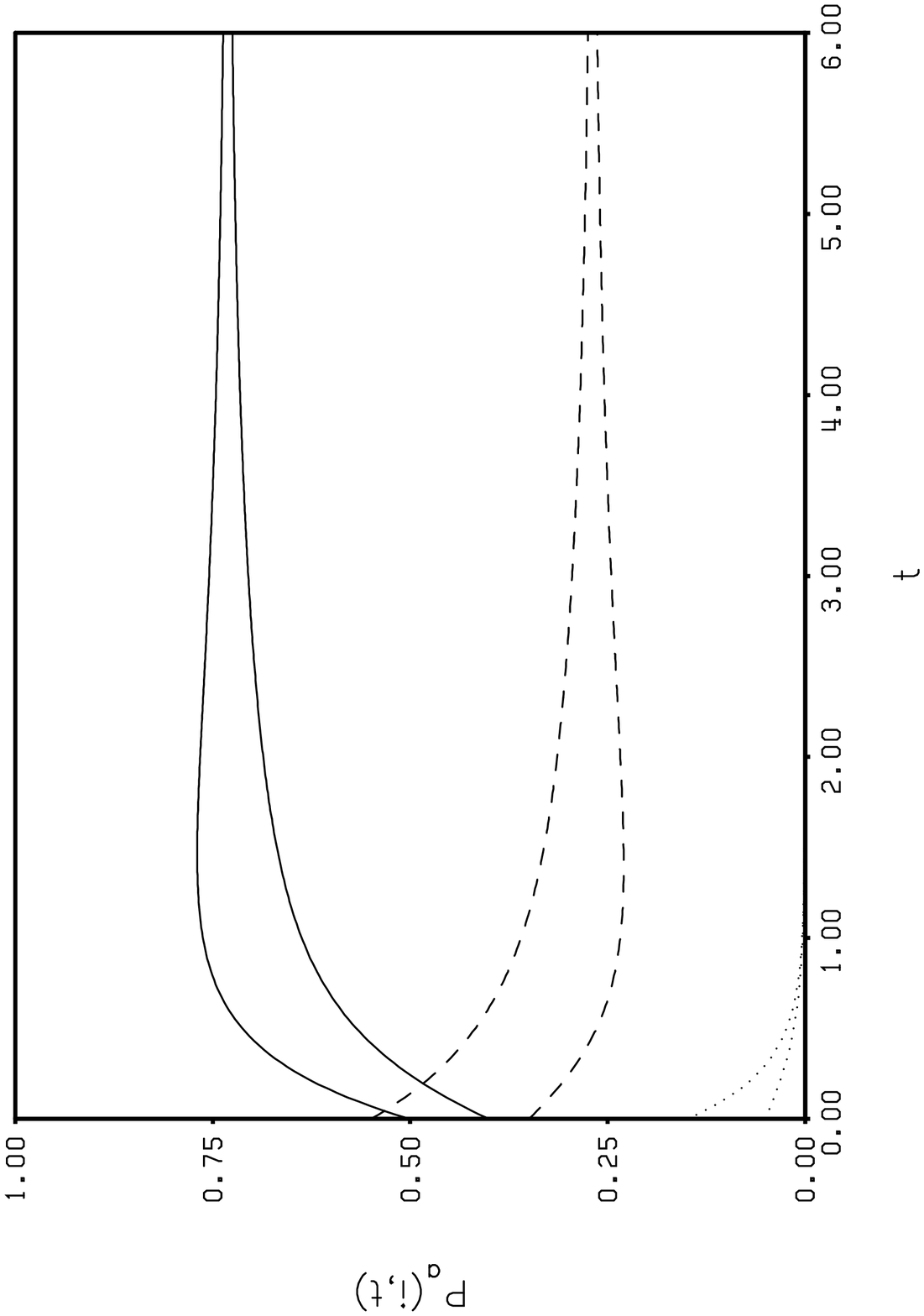}}}}
\capt{Effect of imitative processes for two subpopulations prefering
different behaviors ($c=0.5$): The prefered behavior (---) becomes
the predominating one in each subpopulation, but the behavior which is 
prefered in the {\em other} subpopulation (-- --)
can also convince a certain fraction of individuals.
A behavior which is not prefered by {\em any} subpopulation 
($\cdots$) vanishes.
\label{imit2}}
\end{figure}
\begin{figure}[htbp]
\epsfysize=6.9cm 
\centerline{\rotate[r]{\hbox{\epsffile[57 40 555 756]
{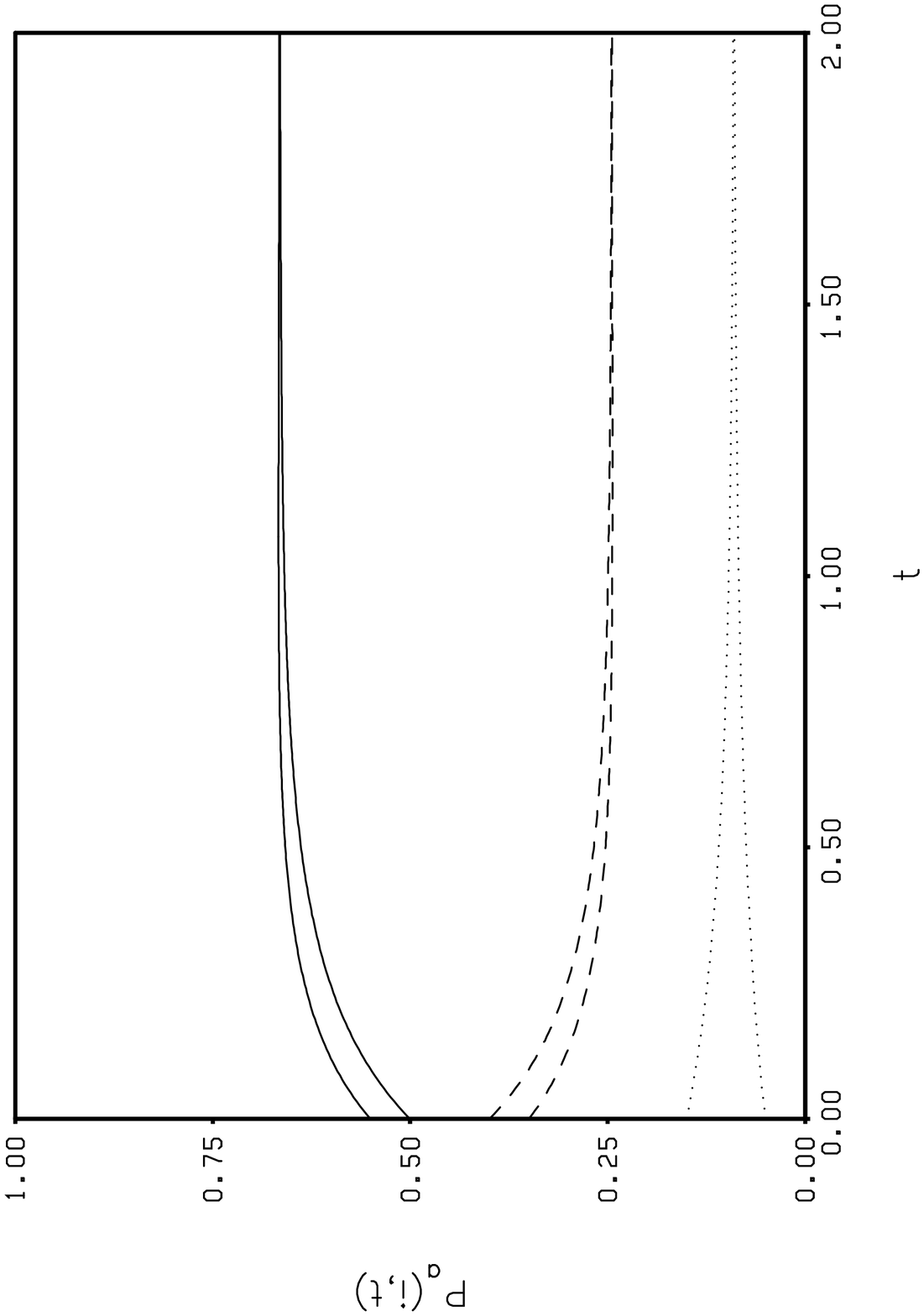}}}}
\capt{Effect of avoidance processes for two subpopulations prefering 
the same behavior ($c=1$): 
The fraction of the prefered behavior (---) is limited, since the
subpopulations avoid to show the same behavior. As a consequence, the other 
behaviors are also used by a certain fraction of individuals.
\label{avoid1}}
\end{figure}
\begin{figure}[htbp]
\epsfysize=6.9cm 
\centerline{\rotate[r]{\hbox{\epsffile[57 40 555 756]
{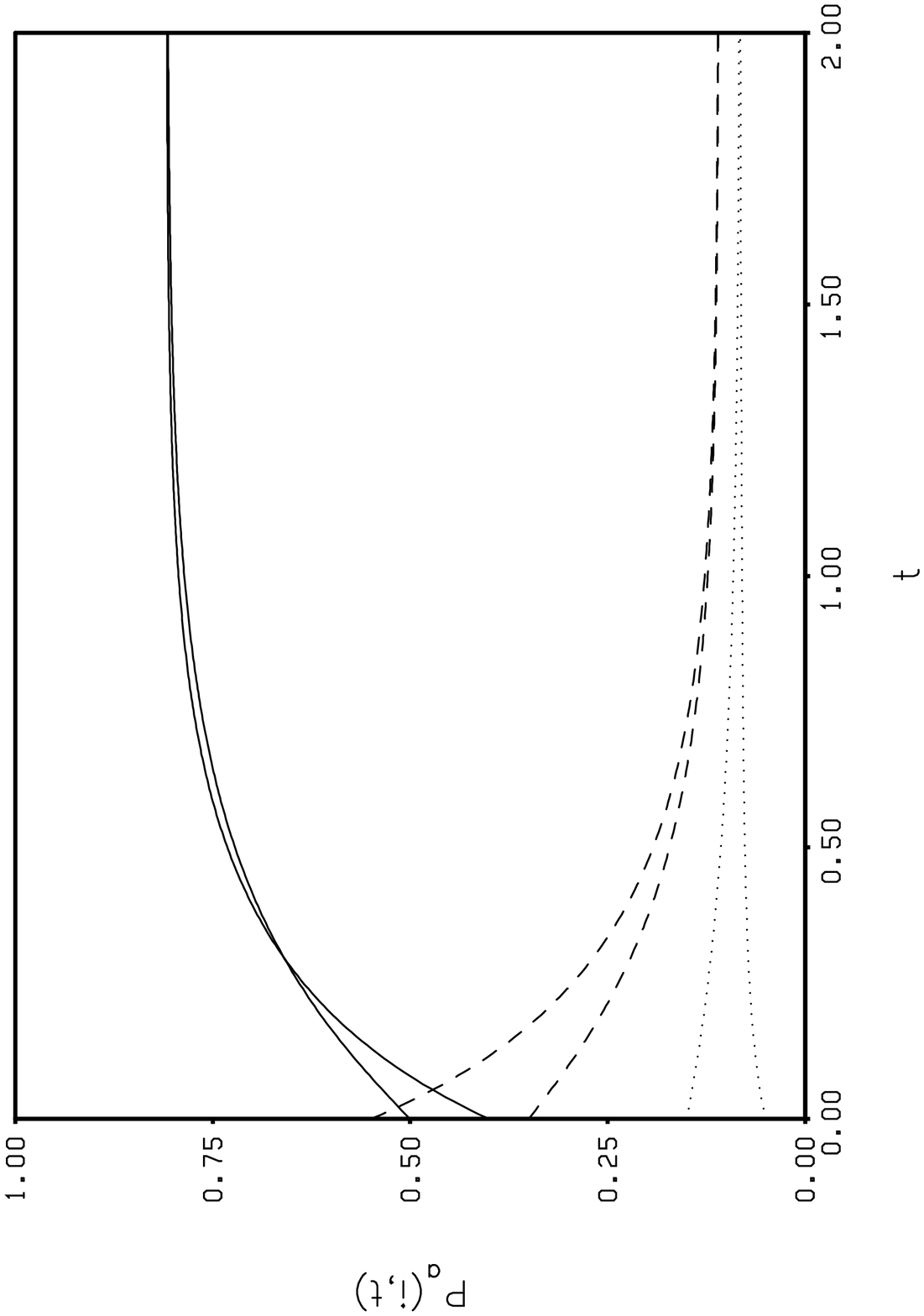}}}} 
\capt{Effect of avoidance processes for two subpopulations prefering 
different behaviors ($c=1$):
The fraction of the prefered behavior (---) wins a greater majority
in comparison with figure \ref{avoid1},
since the situations of avoidance are reduced.
\label{avoid2}}
\end{figure}
\begin{figure}[htbp]
\epsfysize=6.9cm 
\centerline{\rotate[r]{\hbox{\epsffile[57 40 555 756]
{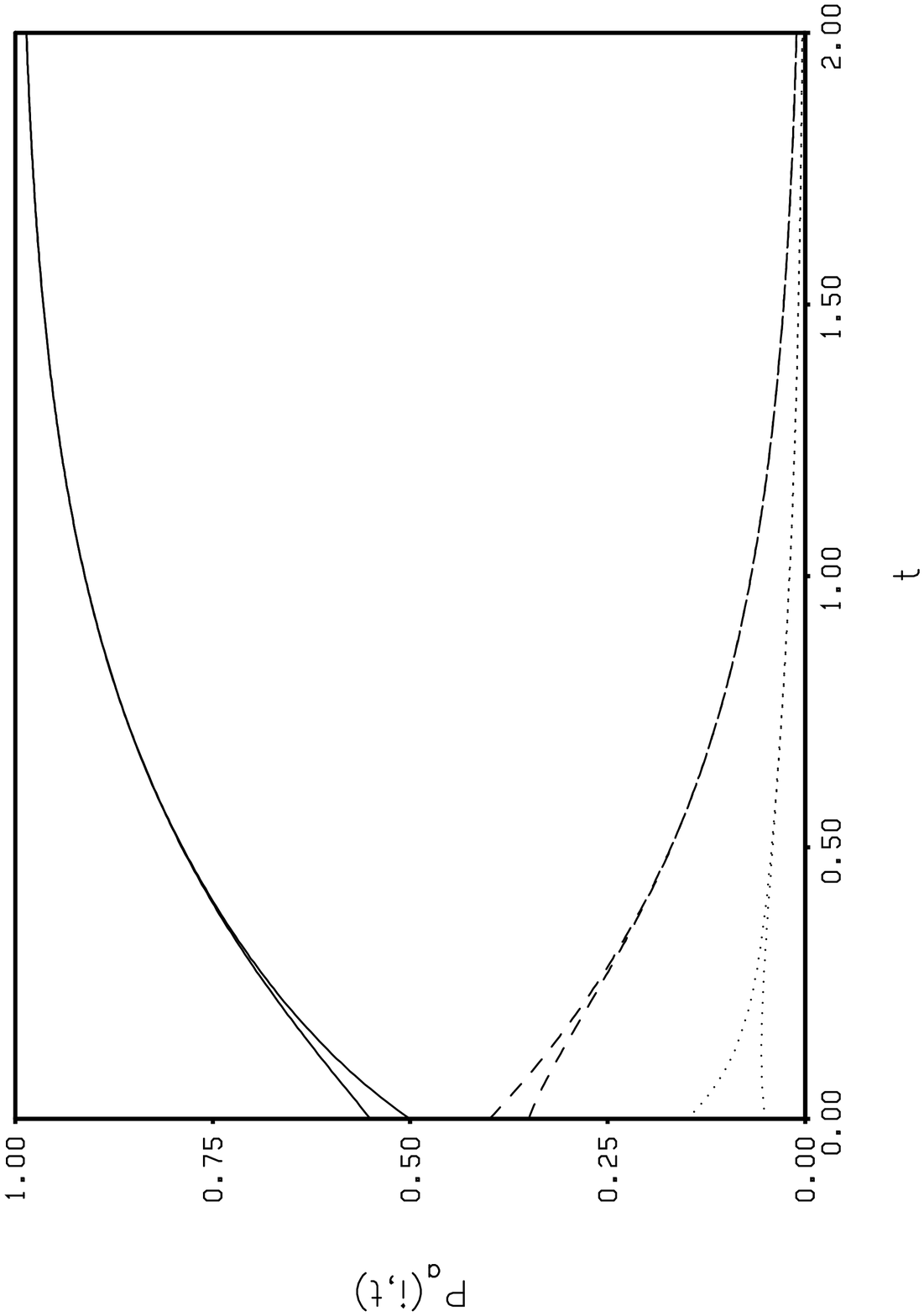}}}}
\capt{Effect of compromising and imitative 
processes for two subpopulations prefering
the same behavior ($c=0.5$):
Only the prefered behavior (---) survives, since a readiness for
compromises is not necessary.
\label{compr1}}
\end{figure}
\begin{figure}[htbp]
\epsfysize=6.9cm 
\centerline{\rotate[r]{\hbox{\epsffile[57 40 555 756]
{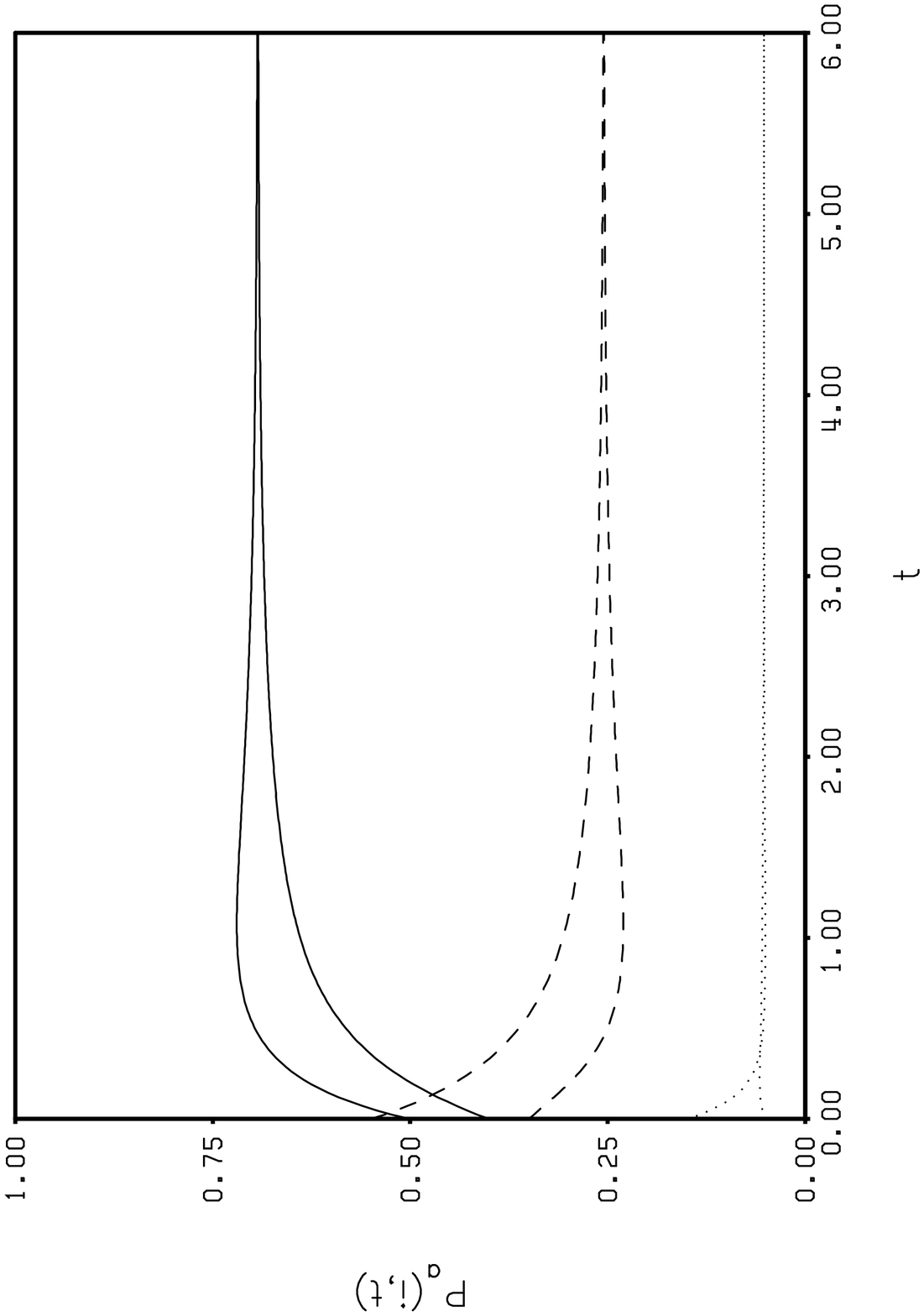}}}}
\capt{Effect of compromising and imitative
processes for two subpopulations prefering
different behaviors ($c=0.5$): Most of the individuals show the 
prefered behavior (---), but a certain
fraction of individuals also decides for a compromise ($\cdots$).
\label{compr2}}
\end{figure}
\begin{figure}[htbp]
\epsfysize=6.9cm 
\centerline{\rotate[r]{\hbox{\epsffile[57 28 555 
756]{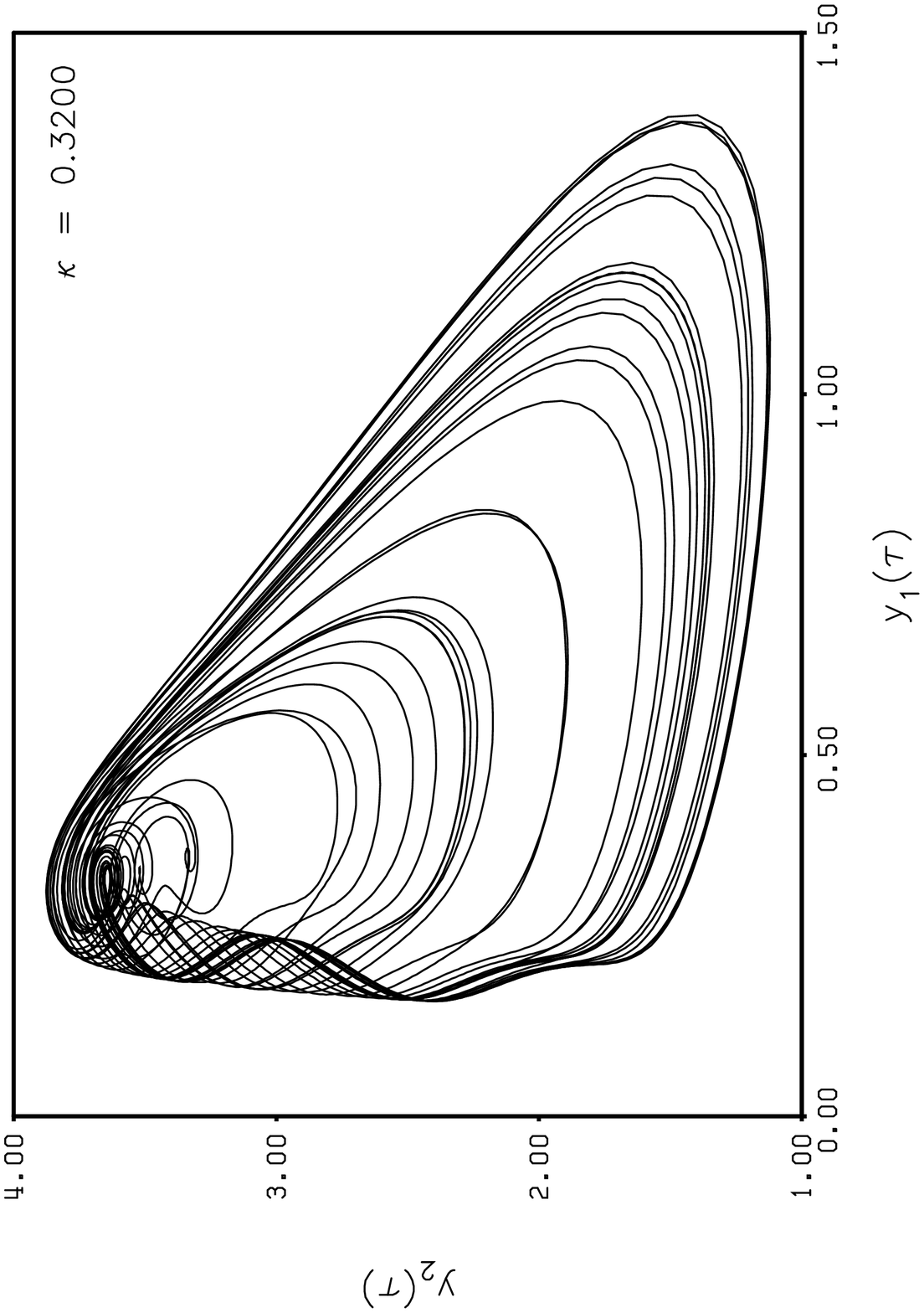}}}}
\capt{Phase portrait of the scaled variables
$y_i(\tau) := \alpha_i P(i,\beta t)$ representing chaotic changes of 
the behavioral fractions $P(i,t)$. 
\label{chaotic}}
\end{figure}

In the following computer simulations $D_a(i',i;t) \equiv 1$ has been taken.
For both subpopulations the {\em prefered} behavior, i.e., 
the behavior with the {\em greatest} utility $U_a(i,t)$
is represented by a solid line, whereas the behavior with the 
lowest utility is represented by a dotted line, 
and the behavior with medium utility by a broken line. 
Figures \ref{imit1} to \ref{compr2} 
show the effects of imitative processes ($\nu_{ab}^1(t) \equiv 1$,
$\nu_{ab}^2(t) \equiv 0 \equiv \nu_{ab}^3(t)$), of avoidance processes
($\nu_{ab}^2(t) \equiv 1$, $\nu_{ab}^1(t) 
\equiv 0 \equiv \nu_{ab}^3(t)$),
resp. of compromising and imitative processes ($\nu_{ab}^3(t) \equiv 1
\equiv \nu_{ab}^1(t)$, $\nu_{ab}^2(t) \equiv 0$)
\begin{itemize}
\item[a)] for {\em equal} behavioral preferenes
($U_1(1) = c=U_2(1)$, $U_1(2)=0=U_2(2)$, 
$U_1(3) = -c=U_2(3)$), and
\item[b)] for {\em different} behavioral preferences
($U_1(1)=c=U_2(2)$, $U_1(2)=0=U_2(1)$, $U_1(3)=-c=U_2(3)$).
\end{itemize}
In more complicated cases, there are also {\em oscillatory}
or {\em chaotic} behavioral changes possible, as illustrated in figures
\ref{oscill} (see {\sc Helbing} (1992b, 1992))
and \ref{chaotic} (see {\sc Helbing} (1992)).
\par
\alphfig{imit}
%
\resetfig
\alphfig{avoid}
\pagebreak
\resetfig
\alphfig{compr}
\resetfig
%
\alphfig{oscill}
\begin{figure}[htbp]
\epsfysize=6.9cm 
\centerline{\rotate[r]{\hbox{\epsffile[57 40 555 756]{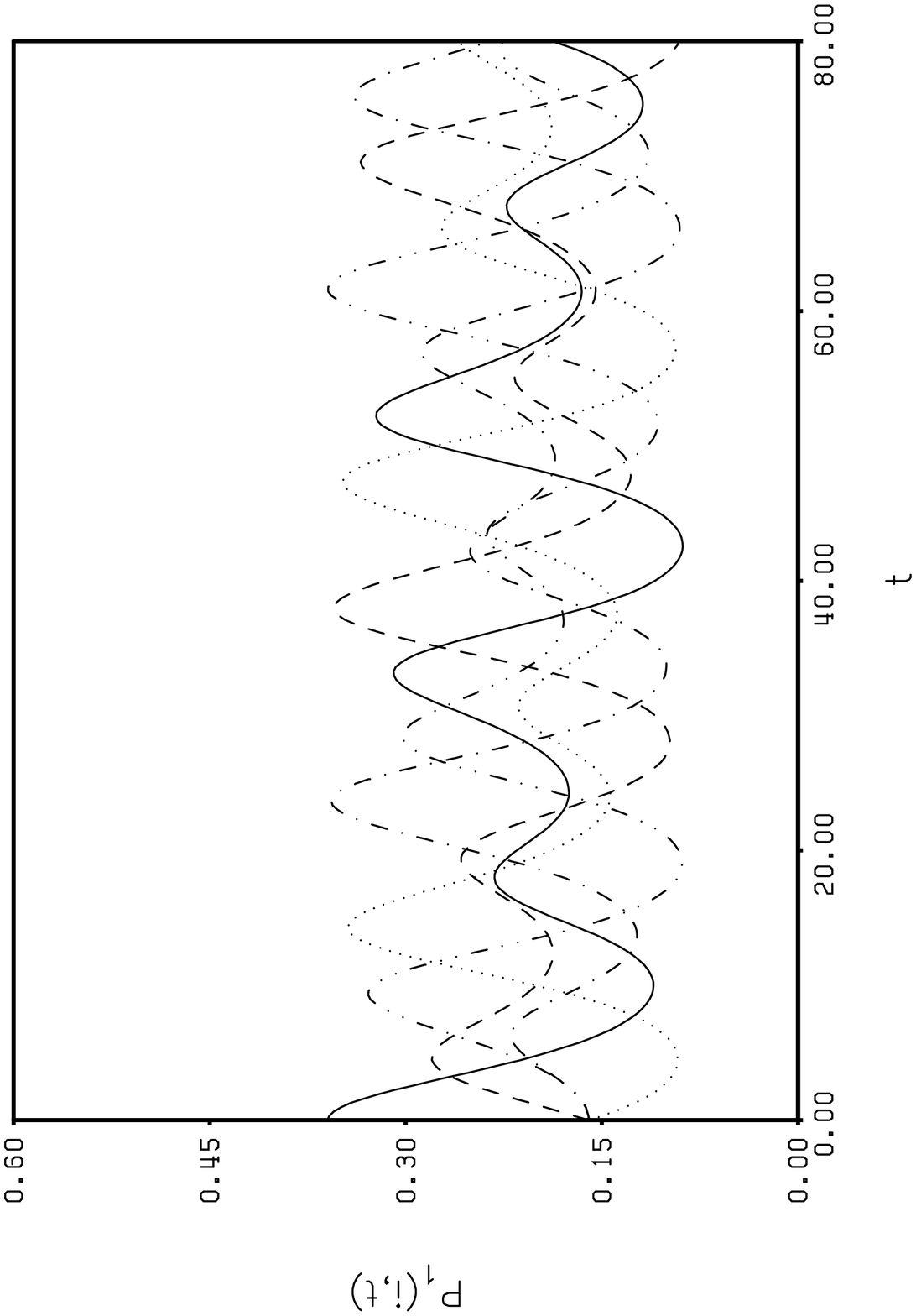}}}}
\capt{Oscillations are one possible effect of imitative 
processes. For $S=5$ different behaviors, the oscillatory changes look
quite irregular without a short-term periodicity.%
\label{oscill1}}
\end{figure}
%
\begin{figure}[htbp]
\epsfysize=6.9cm                   
\centerline{\rotate[r]{\hbox{\epsffile[57 40 555 613]
{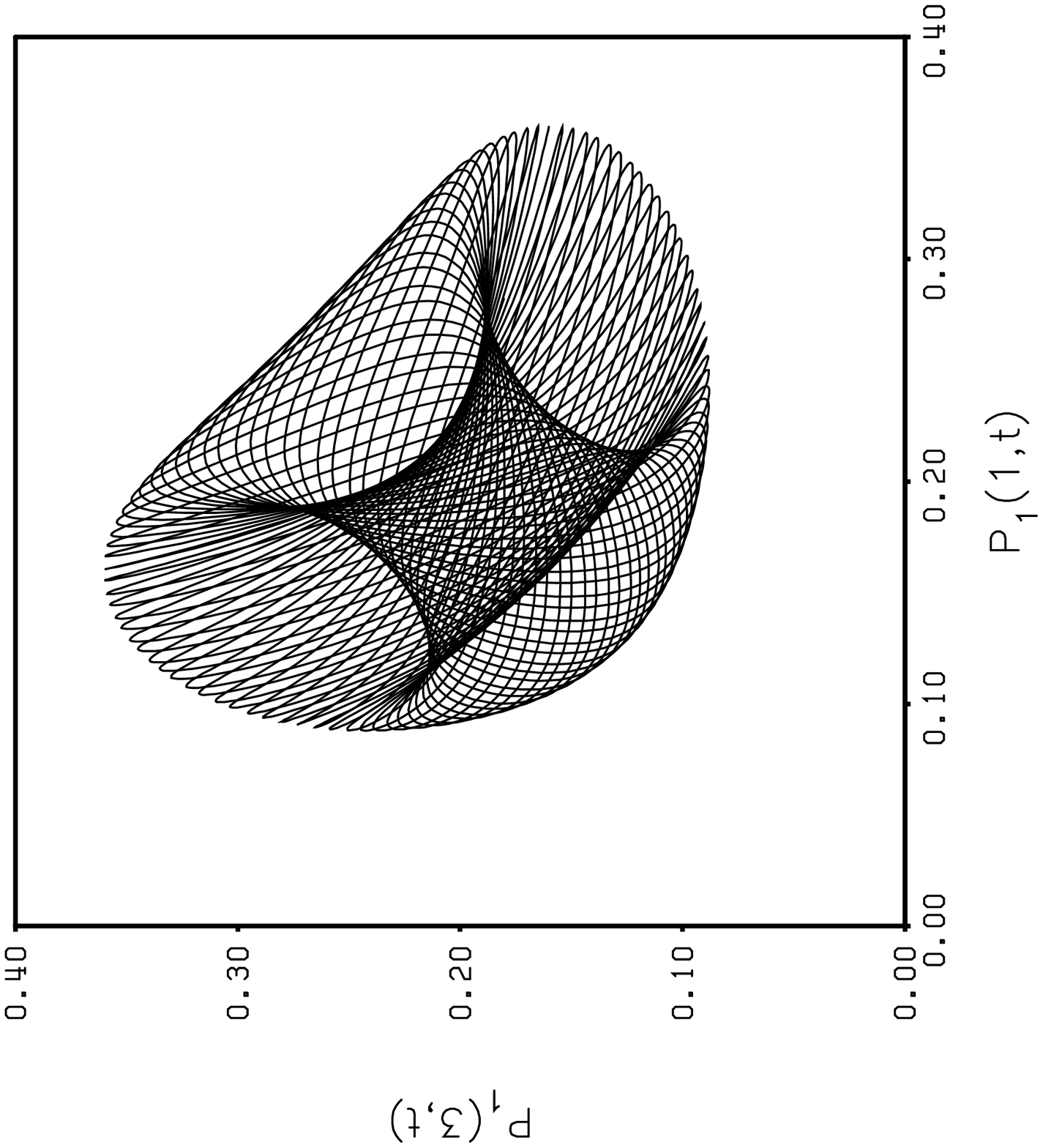}}}}
\capt{Phase portrait of oscillatory changes 
between $S=5$ behaviors having the shape of a torus: A long-term periodicity
is indicated by the closeness of the curve.
\label{oscill2}}
\end{figure}\resetfig\pagebreak
\section{Game dynamical equations} \label{dyngame}

In game theory, $i$ denotes a (behavioral) {\em strategy}. Let
$E_a(i,t)$ be the {\em expected success} of a strategy $i$ for
an individual of subpopulation $a$, and
\begin{displaymath}
 \langle E_a \rangle := \sum_i E_a(i,t)P_a(i,t)
\end{displaymath}
the {\em mean expected success}. If the {\em relative} increase
\begin{displaymath}
 \frac{dP_a(i,t)/dt}{P_a(i,t)}
\end{displaymath}
of the fraction $P_a(i,t)$ is assumed to be proportional to the difference
$[E_a(i,t) - \langle E_a \rangle ]$ between 
the expected and the mean expected success, one obtains
the {\em game dynamical equations}
\begin{equation}
 \frac{d}{dt} P_a(i,t) = \nu_a(t) P_a(i,t) \Big[ E_a(i,t) - \langle
 E_a \rangle \Big] \, .
\label{sel}
\end{equation}
That means, the fractions of strategies with an expected success 
that exceeds the average $\langle E_a \rangle$ are growing, 
whereas the fractions
of the remaining strategies are falling.
For the expected success $E_a(i,t)$, one often takes the form
\begin{equation}
 E_a(i,t) := \sum_b \sum_j A_{ab}(i,j;t)P_b(j,t) \, ,
\label{payoff}
\end{equation}
where $A_{ab}(i,j;t)$ have the meaning of {\em payoffs}. We shall assume
\begin{displaymath}
 A_{ab}(i,j;t) := r_{ab}(t) E_{ab}(i,j;t)
\end{displaymath}
with
\begin{displaymath}
r_{ab}(t) := \frac{\nu_{ab}(t)}{\displaystyle
\sum_c \nu_{ac}(t)} \, ,
\end{displaymath}
where $r_{ab}(t)$ is the {\em relative contact rate} of an individual
of subpopulation $a$ with individuals of subpopulation $b$, and
$E_{ab}(i,j;t)$ is the {\em success} of strategy $i$ for an individual of
subpopulation $a$ during an interaction 
with an individual of subpopulation $b$ who
uses strategy $j$. Since $r_{ab}(t) P_b(j,t)$ is the relative contact rate
of an individual of subpopulation $a$ with individuals of subpopulation $b$
who use strategy $j$, $E_a(i,t)$ is the 
mean (or {\em expected}) success of strategy
$i$ for an individual of subpopulation $a$ in interactions with other
individuals.
\par
By inserting (\ref{payoff}) and
\begin{displaymath}
 \langle E_a \rangle = \sum_{i'} \sum_{b,j} P_a(i',t) A_{ab}(i',j;t)P_b(j,t)
\end{displaymath}
into (\ref{sel}), one obtains the explicit form
\begin{eqnarray}
& & \frac{d}{dt} P_a(i,t) \nonumber \\
&=& \nu_a(t) P_a(i,t) \Bigg[ \sum_{b,j} A_{ab}(i,j;t)
P_b(j,t) \nonumber \\
& & - \; \sum_{i'} \sum_{b,j} P_a(i',t)A_{ab}(i',j;t)P_b(j,t) \Bigg] \nonumber
\\
& &
\label{game}
\end{eqnarray}
of the game dynamical equations. (\ref{game}) is a {\em continuous} formulation
of game theory (see {\sc Hofbauer} and {\sc Sigmund} (1988)). 
Equations of this kind are
very useful for the investigation and understanding of the competition or
cooperation of individuals (see, e.g., {\sc Mueller} (1990),
{\sc Hofbauer} and {\sc Sigmund} (1988),
{\sc Schuster} et. al. (1981)). 
\par
A slightly generalized form of 
(\ref{sel}),\alpheqn{Game}
\begin{eqnarray}
& & \frac{d}{dt} P_a(i,t) \nonumber \\
&=& \sum_{i'} \Big[ w_a(i,i';t)P_a(i',t) \nonumber \\
& & \quad - \; w_a(i',i;t)P_a(i,t) \Big] \label{mutation} \\
&+& \nu_a(t) P_a(i,t) \Big[ E_a(i,t) - \langle E_a \rangle \Big] \, , \qquad 
\label{selection}
\end{eqnarray}\reseteqn
is also known as {\em selection mutation equation} 
({\sc Hofbauer} and {\sc Sigmund} (1988)):
(\ref{selection}) can be understood as effect of a {\em selection}
(if $E_a(i,t)$ is interpreted as {\em fitness} of strategy $i$), and
(\ref{mutation}) can be understood as effect of {\em mutations}. Equation
(\ref{Game}) is a powerful tool in 
evolutionary biology (see {\sc Eigen} (1971), {\sc Fisher} (1930),
{\sc Eigen} and {\sc Schuster} (1979), {\sc Hofbauer} and {\sc Sigmund} 
(1988), {\sc Feistel} and {\sc Ebeling} (1989)).
In game theory, the mutation term could be used for the description
of {\em trial and error} behavior or of accidental variations of the strategy.

\subsection{Connection bet\-ween {\sc Boltz}\-{\sc mann}-like 
and game dy\-nami\-cal equa\-tions}

One expects that there {\em must} be a connection between the {\sc
Boltzmann}-like equations (\ref{eff1}), (\ref{eff2}) and the game dynamical
equations (\ref{Game}), since they 
are both quantitative models for behavioral changes.
A comparison of (\ref{eff1}), (\ref{eff2}) with (\ref{Game}) shows, that both
models can become identical only under the conditions
\begin{eqnarray}                              
 \nu_{ab}^1(t) &=& \nu_a(t) \delta_{ab}\, , \nonumber \\
 \nu_{ab}^2(t) &=& 0\, , \nonumber \\
 \nu_{ab}^3(t) &=& 0\, .
\label{nu}
\end{eqnarray}
That means, the game dynamical equations include spontaneous and imitative
behavioral changes, but they exlude avoidance and compromising processes.
\par
In order to make the analogy between the game dynamical and the {\sc
Boltzmann}-like equations complete the following assumptions have to be made:
\begin{itemize}
\item In interactions with other individuals the expected success 
\begin{eqnarray}
 & & E_a(i,t) \nonumber \\
&=& \sum_{b,j} \frac{\nu_{ab}(t)}{\displaystyle \sum_c \nu_{ac}(t)}
 E_{ab}(i,j;t) P_b(j,t) \nonumber \\
& &
\label{suc}
\end{eqnarray}
of a strategy is evaluated. This is possible, since an individual is able to
determine the quantities $\nu_{ab}(t)$, $P_b(j,t)$ and $E_{ab}(i,j;t)$:
An individual of subpopulation $a$ meets individuals of
subpopulation $b$ with a contact rate of $\nu_{ab}(t)$. With a probability
of $P_b(j,t)$, the individuals of subpopulation $b$ use the strategy
$j$. During interactions with individuals of subpopulation $b$ who use the
strategy $j$, an individual of
subpopulation $a$ has a success of $E_{ab}(i,j;t)$ if using the strategy
$i$. 
\item In interactions with individuals of the 
{\em same} subpopulation an individual
tends to take over the strategy of another individual, if the expected success
would increase: 
If an individual who uses a strategy $i$ meets another individual
of the same subpopulation who uses a strategy $j$, they will compare their
expected success' $E_a(i,t)$ resp. $E_a(j,t)$ (by observation or exchange
of their experiences). The individual with strategy $i$ will imitate
the other's strategy $j$ with a probability $p_{ab}^1(j|i;t)$ that is 
growing with the expected increase
\begin{displaymath}
 \Delta_{ji} E_a := E_a(j,t) - E_a(i,t)
\end{displaymath}
of success. If a change of strategy would imply a decrease of success
($\Delta_{ji} E_a < 0$), the individual will not change the strategy $i$.
Therefore, the readiness for replacing the strategy $i$ by $j$ during an
interaction within the same subpopulation can be assumed to be
\begin{equation}
 R_a(j,i;t) := \max \Big( E_a(j,t) - E_a(i,t),0 \Big) \, ,
\label{read}
\end{equation}
where $\max(x,y)$ is the maximum of the two numbers $x$ and $y$.
However, due to different criteria for the grade of success, the expected
success of a strategy $i$ will usually be varying with the subpopulation $a$
(i.e., $E_a(i,t) \ne E_b(i,t)$ for \mbox{$a\ne b$}). As a consequence,
an imitative behavior of individuals belonging to 
{\em different} subpopulations is
not plausible, and we shall assume
\begin{displaymath}
 f_{ab}^1(t) := \delta_{ab} \, ,
\end{displaymath}
that means,
\begin{displaymath}
 \nu_{ab}^1(t) = \nu_{aa}(t) \delta_{ab} \, .
\end{displaymath}
\end{itemize}
Inserting (\ref{nu}), (\ref{suc}) and (\ref{read}) into the 
{\sc Boltzmann}-like equations (\ref{eff1}), (\ref{eff2}), the game dynamical
equations (\ref{Game}) result as a special case, since
\begin{eqnarray*}
& & \max \Big( E_a(i,t)-E_a(j,t),0 \Big) \nonumber \\
&-& \max \Big( E_a(j,t) - E_a(i,t),0 \Big) \nonumber \\
&=& E_a(i,t) - E_a(j,t) \, .
\end{eqnarray*}

\subsection{Stochastic version of the game dynamical equations} 
\label{stochgame}
Applying the formalism of section \ref{stoch}, a stochastic version of the
game dynamical equations can easily be formulated. This is given by the
master equation (\ref{master}) with the configurational transition rates
(\ref{rate}) and
\begin{eqnarray*}
& & w_{ab}(i',j';i,j;t) \nonumber \\
&:=& N_b \widetilde{w}_{ab}^1(i',j';i,j;t) \nonumber \\
&:=& \nu_a(t) \delta_{ab}
 \widehat{R}_{a}(i,j;t)\delta_{ii'}\delta_{ij'}(1-\delta_{ij}) \nonumber \\
&+& \nu_a(t) \delta_{ab} \widehat{R}_{a}(j,i;t) 
\delta_{jj'} \delta_{ji'}(1-\delta_{ij}) \, ,
\end{eqnarray*}
where                                                
\begin{displaymath}
 \widehat{R}_a(j,i;t) 
 := \max \Big( \widehat{E}_a(j,t) - \widehat{E}_a(i,t), 0 \Big)
\end{displaymath}
and 
\begin{displaymath}
 \widehat{E}_a(i,t) := \sum_b \sum_j A_{ab}(i,j;t) \frac{n_j^b}{N_b} 
\end{displaymath}
(compare to {\sc Feistel} and {\sc Ebeling} (1989), 
{\sc Ebeling} and {\sc Feistel} (1982), {\sc Ebeling} et. al. (1990)).
A comparison with (\ref{Mean}), (\ref{eff1}), (\ref{eff2})
shows, that the ordinary game dynamical equations
(\ref{Game}) are the {\em approximate} mean value equations of this special
master equation. Therefore, they can only be interpreted as mean
value equations as long as the (co)variances $\sigma_{ij}^{ab}$
are small (see (\ref{sm})). Otherwise they describe the most probable
behavioral distribution (see sect. \ref{Most}).

\subsection{Self\-or\-ga\-ni\-za\-tion of be\-ha\-vio\-ral conventions by
com\-pe\-ti\-tion between equi\-va\-lent stra\-te\-gies} \label{beisp}

This section gives an illustration of the methods and results derived in
sections \ref{dyngame} and \ref{stochgame}. 
As an example, we shall consider a case
with one subpopulation only ($A=1$), and, therefore, omit the index $a$
in the following. Let us suppose the individuals to choose between two
equivalent strategies $i \in \{1,2\}$, i.e., the payoff matrix 
$\underline{A}(t)$ shall be symmetrical:
\begin{equation}
 \underline{A}(t) \equiv \Big( A(i,j;t) \Big) :=
\left(
\begin{array}{cc}
A+B & B \\
B & A+B
\end{array}\right) \, .
\label{pay}
\end{equation}
According to the relation 
\begin{displaymath}
 n_1 + n_2 = N 
\end{displaymath}
(see (\ref{sum})),
the fraction $P(2,t) = 1 - P(1,t)$ is already determined by $P(1,t)$.
By scaling the time, 
\begin{displaymath}
\nu(t) \equiv 1
\end{displaymath}
can be presupposed. For the spontaneous change of strategies due to trial and 
error we shall assume the transition rates
\begin{equation}
 w(j,i;t) := W \, .
\label{fluct}
\end{equation}
A situation of the above kind is the avoidance behavior of pedestrians (see
{\sc Helbing} (1991)): 
In pedestrian crowds with two opposite directions of movement, the
pedestrians have sometimes to avoid each other in order to exclude a collision.
%
%
For an avoidance maneuver to be successful, both pedestrians concerned have to
pass the respective other pedestrian either on the right hand side or on
the left hand side. Otherwise, both pedestrians have to stop 
(see figure \ref{separation}).
Therefore, both strategies 
(to pass pedestrians on the right hand side 
or to pass them on the left hand side)
are equivalent, but the success of a strategy grows
with the number $n_i$ of individuals who use the {\em same} strategy.
In the payoff matrix (\ref{pay}) we have $A>0$, then.
\par
\begin{figure}[htbp]
\unitlength1cm
\begin{center}
\begin{picture}(3.4,3)(0,1.5)
\thinlines
\put(1.7,1.8){\circle*{0.54}}
\put(1.8,2.1){\vector(1,2){0.9}}
\dashline{0.2}(1.6,2.1)(0.7,3.83)
\put(0.7,3.83){\vector(-1,2){0}}
\put(1.7,4.2){\circle{0.54}}
\put(1.6,3.9){\vector(-1,-2){0.9}}
\dashline{0.2}(1.8,3.9)(2.7,2.17)
\put(2.7,2.17){\vector(1,-2){0}}
\put(0.2,2.2){\makebox(0,0){$P(1)$}}
\put(3.2,2.2){\makebox(0,0){$P(2)$}}
\put(0.2,3.8){\makebox(0,0){$P(2)$}}
\put(3.2,3.8){\makebox(0,0){$P(1)$}}
\end{picture}
\end{center}
\capt{For pedestrians with an opposite direction 
of motion it is advantageous,
if both prefer either the right hand side or the left hand side when
trying to pass each other. Otherwise, they would have to stop in order
to avoid a collision. The probability $P(1)$ for choosing the right hand side
is usually greater than the probability $P(2)=1-P(1)$ 
for choosing the left hand side.\label{separation}}
\end{figure}
%
The game dynamical equations (\ref{Game}) corresponding
to (\ref{pay}), (\ref{fluct}) have the explicit form
\begin{eqnarray}
 \frac{d}{dt} P(i,t) &=& -2\left( P(i,t) - \frac{1}{2} \right) \nonumber \\
&\times& \Big[ W + AP(i,t) \Big( P(i,t) - 1 \Big) \Big] \, . \nonumber \\
& &
\label{concr}
\end{eqnarray}
According to (\ref{concr}), $P(i) = 1/2$
is a stationary solution. This solution is stable only for
\begin{displaymath}
 \kappa := 1 - \frac{4W}{A} < 0 \, ,
\end{displaymath}
i.e., if spontaneous strategy changes due to trial and error 
(the ``mutations'') are dominating.
For $\kappa > 0$ the stationary solution $P(i) = 1/2$ is unstable, and
the game dynamical equations (\ref{concr}) can be rewritten in the form
\begin{eqnarray*}
 \frac{d}{dt} P(i,t) &=& -2 \left( P(i,t) - \frac{1}{2} \right) \nonumber \\
&\times& \left( P(i,t) - \frac{1 + \sqrt{\kappa}}{2} \right) \nonumber \\
&\times& \left( P(i,t) - \frac{1-\sqrt{\kappa}}{2} \right) \, .
\end{eqnarray*}
That means, for $\kappa > 0$ we have two additional stationary solutions 
$P(i) = (1+\sqrt{\kappa})/2$ and $P(i) = (1-\sqrt{\kappa})/2$, which
are stable. Depending on the random initial condition $P(i,t_0)$, one strategy
will win a majority of $100\cdot\sqrt{\kappa}$ 
percent. This majority is the greater,
the smaller the rate $W$ of spontaneous strategy changes is.
\par
\begin{figure}[htbp]
\epsfysize=7.2cm 
\centerline{\rotate[r]{\hbox{\epsffile[28 28 570
556]{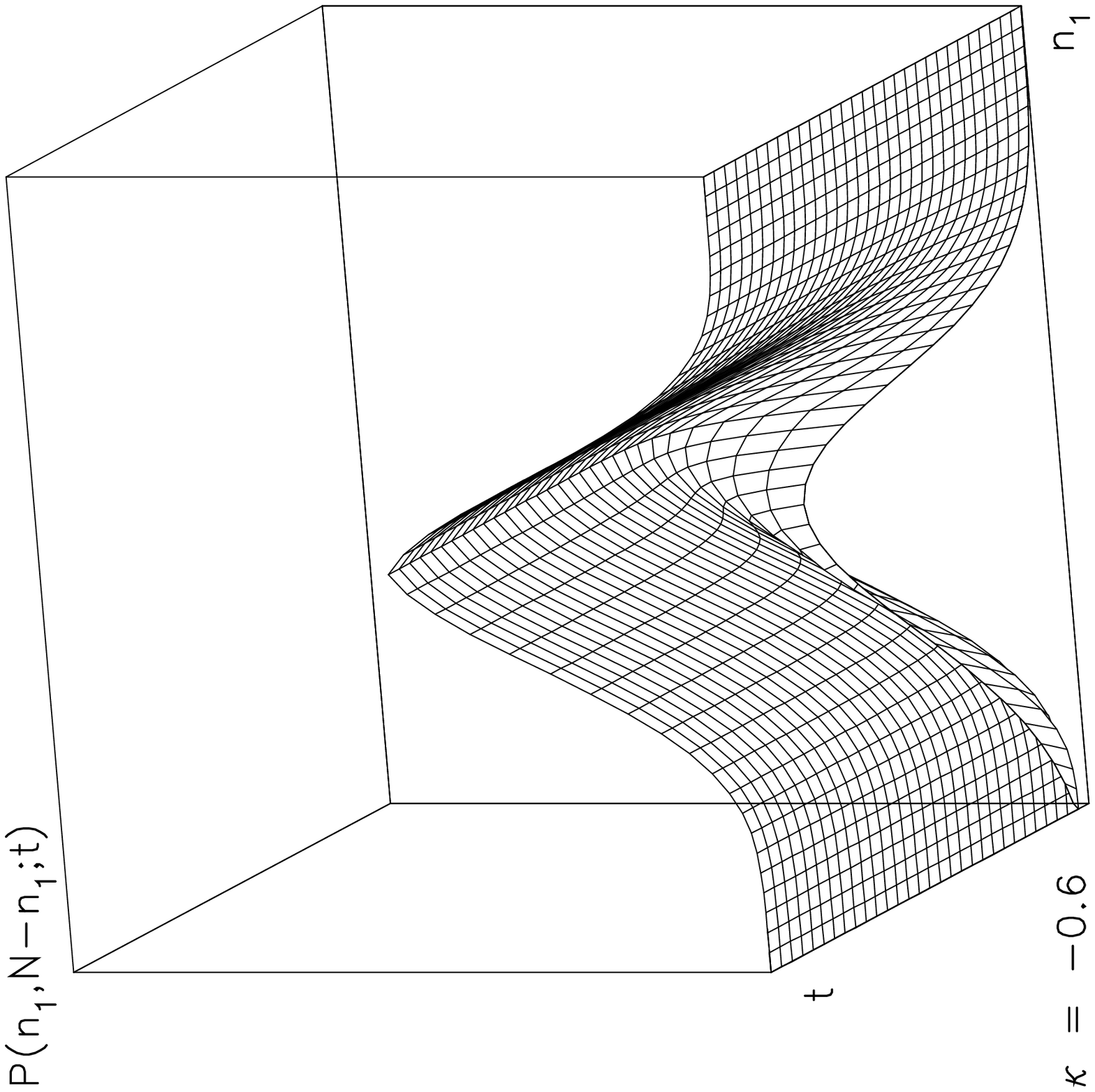}}}}
\capt{Probability distri\-bu\-tion $P(\vec{n},t) \equiv$\linebreak
$P\,(\,n_1\,,\,N-n_1\,;\,t\,)$
of the socio\-con\-fi\-gu\-ra\-tion\linebreak 
$\vec{n}=(n_1,N-n_1)$ for two equi\-va\-lent com\-pe\-ting
strat\-egies. Muta\-tion dominated region ($\kappa < 0$): Since 
$P(n_1,N-n_1;t)$ has, after a certain time interval, 
one maximimum at $n_1 = N/2$, each strategy will
most probably be used by about one half of the individuals.
\label{below1}}
\end{figure}
\begin{figure}[htbp]
\epsfysize=7.2cm 
\centerline{\rotate[r]{\hbox{\epsffile[28 28 570
556]{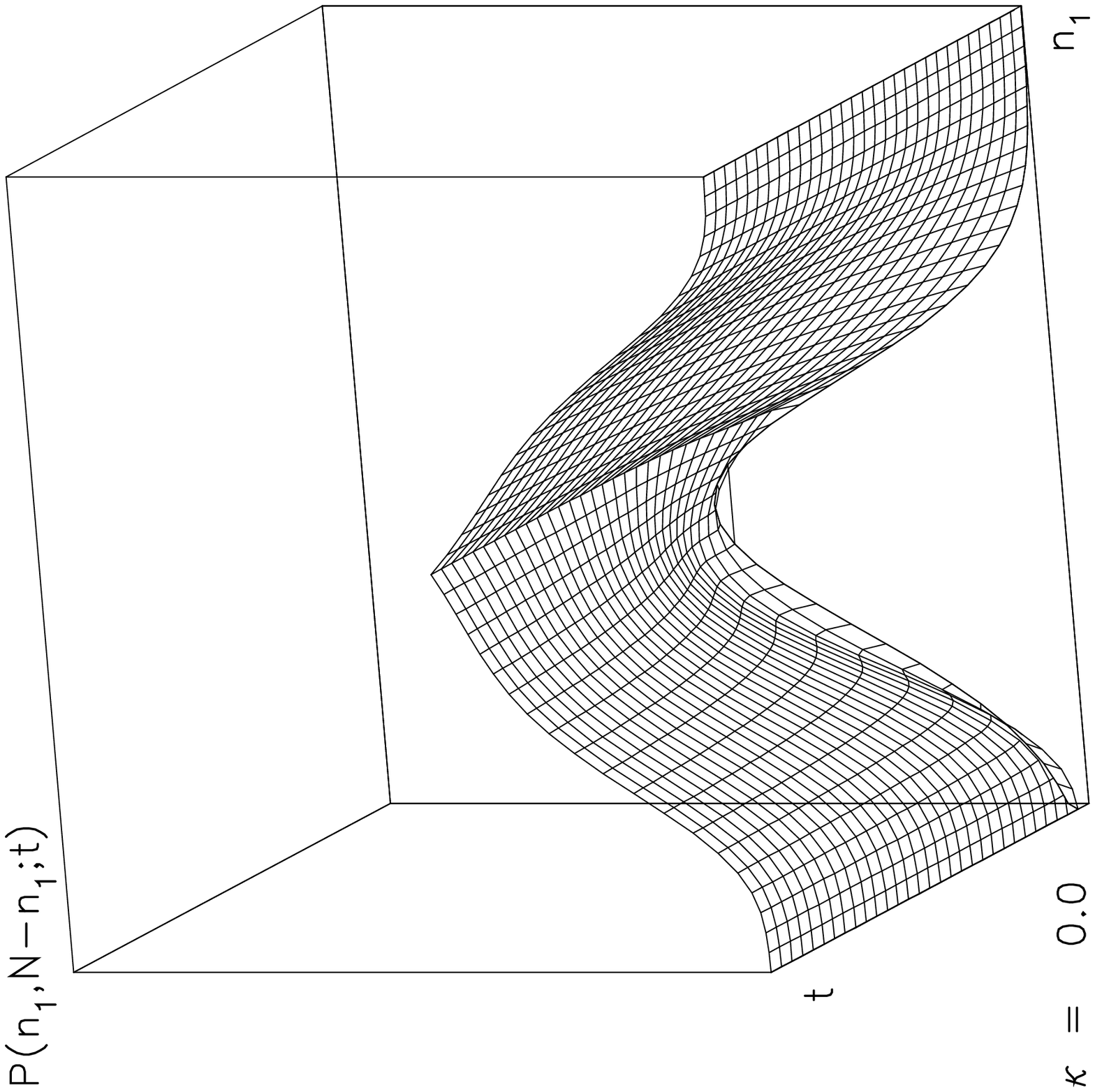}}}}
\capt{As figure \ref{below1}, but for the critical point 
$\kappa=0$: The broadness of the probability distribution 
$P(n_1,N-n_1;t)$ indicates {\em critical
fluctuations}, i.e., a phase transition.
\label{at1}}
\end{figure}
\begin{figure}[htbp]
\epsfysize=7.2cm 
\centerline{\rotate[r]{\hbox{\epsffile[28 28 570
556]{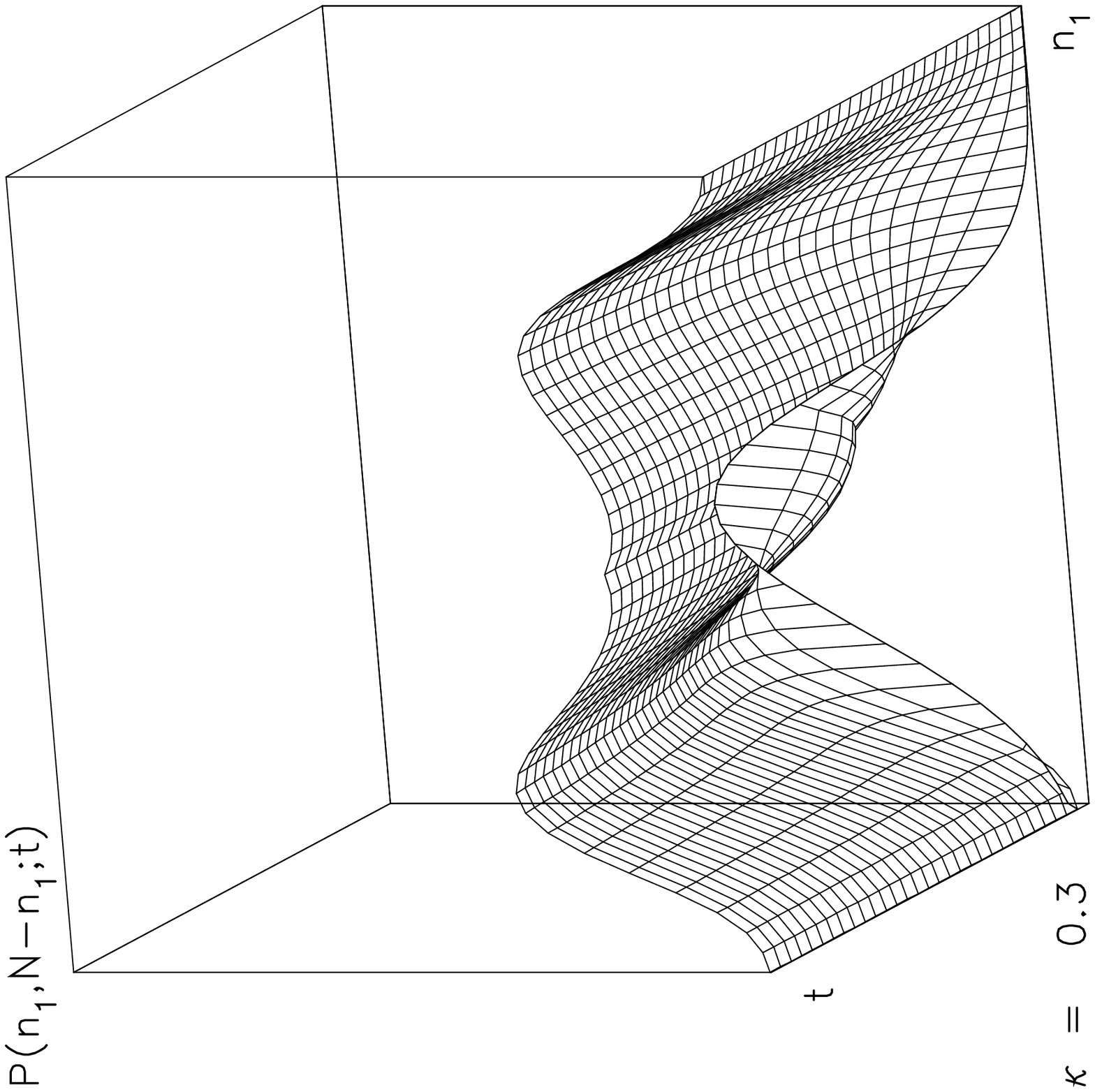}}}}
\capt{As figure \ref{below1}, but after the phase transition
($\kappa > 0$): The configurational distribution $P(n_1,N-n_1;t)$
becomes multimodal with maxima that 
are symmetrical with respect to $N/2$, because of
the equivalence of the strategies. Due to the maxima at $n_1 > N/2$ and
$n_2= N-n_1 > N/2$,
one of the strategies will very probably win a majority of users.
This implies the selforganization of a
behavioral convention.
\label{above1}}
\end{figure}

At the {\em critical point} $\kappa = \kappa_0 := 0$ 
there appears a {\em phase transition}. This can
be seen best in figures \ref{below1} to \ref{above1}, where the distribution
$P(\vec{n},t) \equiv P(n_1,n_2;t) = P(n_1, N-n_1;t)$ loses its unimodal
form for $\kappa > 0$. As a consequence of the phase transition, 
one strategy is prefered, i.e. a behavioral convention develops.
\pagebreak
%
\vfill
%
\pagebreak
%
\vfill
\begin{figure}[htbp]
\epsfysize=7.2cm 
\centerline{\rotate[r]{\hbox{\epsffile[28 28 570
556]{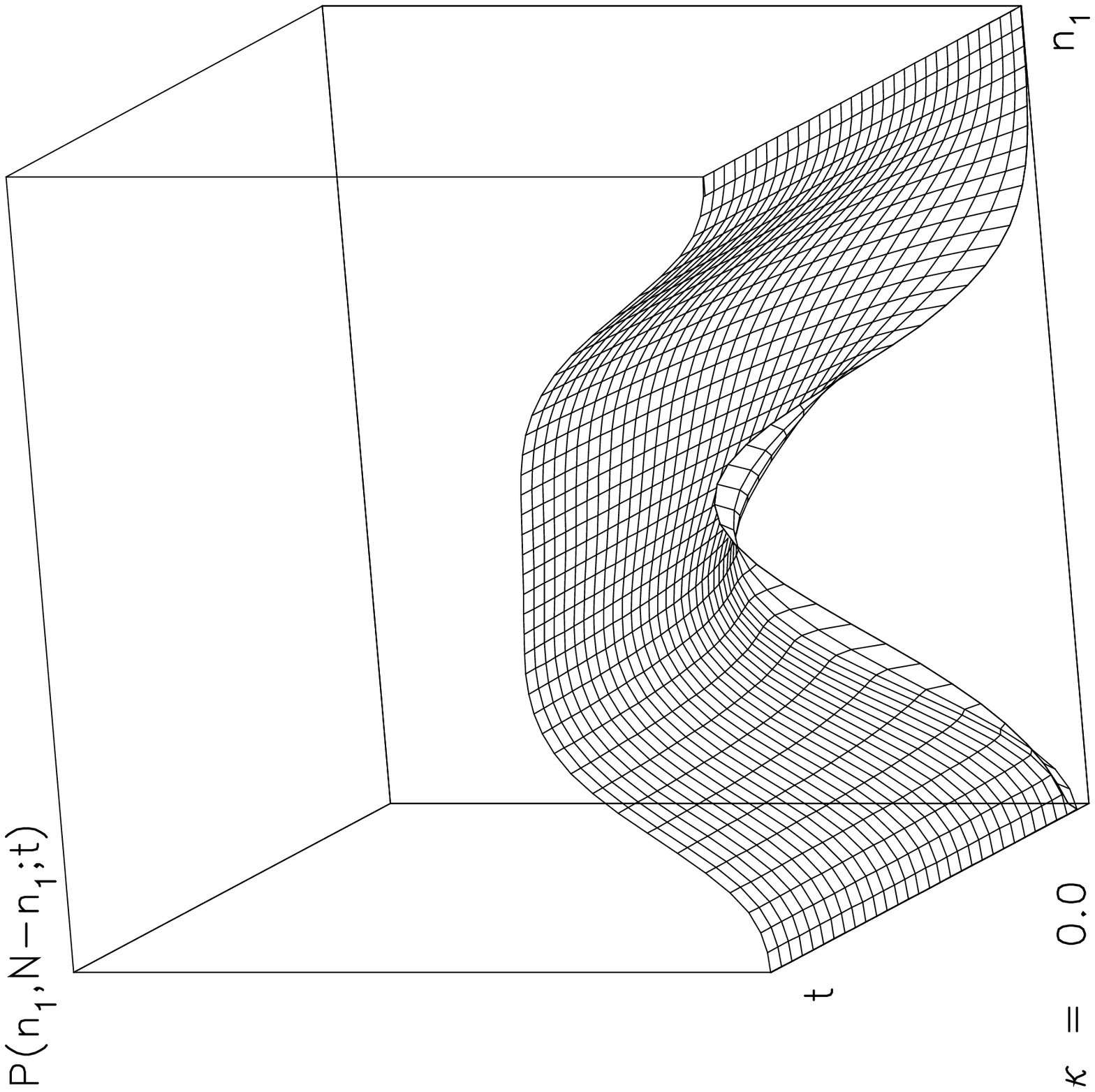}}}}
\capt{As figure \ref{at1}, but with a modified {\em ansatz} for the
readiness $R_a(j,i;t)$ to change the behavior from $i$ to $j$, which does
not produce a crease of $P(n_1,N-n_1;t)$ at $N/2$.
\label{crit}}
\end{figure}
\pagebreak
The crease of $P(n_1,N-n_1;t)$ at 
$n_1 = N/2 = n_2$ is a consequence of the crease of the function
$\widehat{R}_a(j,i;t) = \max 
(\widehat{E}_a(j,t) - \widehat{E}_a(i,t),0)$. 
It can be avoided by using the modified {\em ansatz}
\begin{displaymath}
 \widehat{R}_a(j,i;t) 
:= \frac{\mbox{e}^{\widehat{E}_a(j,t) - \widehat{E}_a(i,t)}}{D_a(j,i;t)}
\end{displaymath}
(compare to (\ref{readiness})), which also shows a phase transition
for $\kappa = 0$
(see figure \ref{crit}).
%

\section{Summary and Conclusions}

A quite general model for behavioral changes has been developed, which takes
into account spontaneous changes and changes due to pair interactions.
Three kinds of pair interactions have been distinguished:
imitative, avoidance and compromising processes.
The game dynamical equations result for a special case of imitative processes.
They can be interpreted as equations for the most probable behavioral
distribution or as approximate mean value equations 
of a stochastic formulation of
game theory. In order to determine the reliability (or representativity) of 
game dynamical descriptions, 
one has to evaluate the corresponding (co)variance equations.

\section*{Acknowledgements}
The author wants to thank Prof. Dr. W. Weid\-lich and
PD Dr. G. Haag for inspiring discussions.
\vspace*{2mm}
\begin{center}
{\large\it References}
\end{center}
\mbox{ }\\[-0.8cm]
\bibitem{Axelrod}{\em Axelrod, R.} (1984), The Evolution of Cooperation.
 New York: Basic Books
\bibitem{Boltzmann}{\em Boltzmann, L.} (1964), Lectures on Gas Theory.
 Berkeley: University of California
\bibitem{Engel}{\em Ebeling, W.\,/\,\,Engel, A.\,/\,\,Feistel, R.}
 (1990),\linebreak Phy\-sik der
 Evo\-lu\-tions\-pro\-zes\-se. Berlin: Aka\-de\-mie-Verlag
\bibitem{Ebeling}{\em Ebeling, W.\,/\,\,Feistel, R.} (1982), Physik der
 Selbstorganisation und Evolution. Berlin: Akademie-Verlag
\bibitem{Eigen}{\em Eigen, M.} (1971), The selforganization of matter 
 and the evolution
 of biological macromolecules. Naturwissenschaften {\bf 58}, 465
\bibitem{Eig}{\em Eigen, M.\,/\,\,Schuster, P.} (1979), The Hypercycle.
 Berlin: Springer
\bibitem{Feistel}{\em Feistel, R.\,/\,\,Ebeling, W.} (1989), Evolution of Complex
 Systems. Dordrecht: Kluwer Academic 
\bibitem{Fisher}{\em Fisher, R. A.} (1930), The Genetical Theory of Natural 
 Selection. Oxford: Oxford University
\bibitem{Goodwin}{\em Goodwin, R. M.} (1969), A growth cycle. 
 In: {\em Feinstein, C. H.} (ed.),
 Socialism, Capitalism and Economic Growth. Cambridge:
 Cambridge University Press.
 Revised version in: {\em Hunt, E. K.\,/\,\,Schwarz, J. G.} (eds.), A Critique
 of Economic Theory. Harmondsworth: Penguin, pp. 442-449
\bibitem{Goel}{\em Goel, N. S.\,/\,\,Maitra, S. C.\,/\,\,Montroll, E. W.} (1971),
 Reviews of Modern Physics {\bf 43}, 231-276
\bibitem{Haken}{\em Haken, H.} (1983), Synergetics. An Introduction.
 Berlin: Springer, pp. 79-83
\bibitem{Hallam}{\em Hallam, Th. G.} (1986), Community dynamics in a homogeneous
 environment. In: {\em Hallam, Th. G.\,/\,\,Levin, S. A.} (eds.) 
 Mathematical Ecology. Berlin: Springer, pp. 241-285
\bibitem{Helbing}{\em Helbing, D.} (1991), A mathematical model for the behavior of
 pedestrians. Behavioral Science {\bf 36}, 298-310
\bibitem{Hel}{\em Helbing, D.} (1992), Stochastische Methoden, nichtlineare
Dynamik und quantitative Modelle sozialer Prozesse. 
Universit\"at Stuttgart: Dissertation
\bibitem{Helbinga}{\em Helbing, D.} (1992a), Interrelations between stochastic equations
 for systems with pair interactions. Physica A {\bf 181}, 29-52
\bibitem{Helbingb}{\em Helbing, D.} (1992b), A mathematical model for attitude formation 
 by pair interactions. Behavioral Science {\bf 37}, 190-214
\bibitem{Hof}{\em Hofbauer, J.} (1981), On the occurence of limit cycles in the
 Lotka-Volterra equation. Nonlinear Analysis {\bf TMA 5},
 1003-1007
\bibitem{Hofbauer}{\em Hofbauer, J.\,/\,\,Sigmund, K.} (1988), The Theory of Evolution
 and Dynamical Systems. Cambridge: Cambridge University Press
\bibitem{Lotka}{\em Lotka, A. J.} (1920), Proc. Nat. Acad. Sci. U.S. 
 {\bf 6}, 410
\bibitem{Lotka1}{\em Lotka, A. J.} (1956), Elements of Mathematical Biology.
 New York: Dover
\bibitem{Luce}{\em Luce, R. D.\,/\,\,Raiffa, H.} (1957), Games and Decisions.
 New York: Wiley
\bibitem{Mueller}{\em Mueller, U.} (ed.) (1990), Evolution und Spieltheorie.
 M\"unchen: Oldenbourg
\bibitem{Neumann}{\em von Neumann, J.\,/\,\,Morgenstern, O.} (1944), Theory of Games
 and Economic Behavior. Princeton: Princeton University Press
\bibitem{Pearl}{\em Pearl, R.} (1924), Studies in Human Biology. Baltimore:
 Williams \& Wilkins
\bibitem{Rav}{\em Ravenstein, E.} (1876), The birthplaces of the people and the laws
 of migration. The Geographical
 Magazine {\bf III}, 173-177, 201-206, 229-233
\bibitem{Schust}{\em Schuster, P.\,/\,\,Sigmund, K.\,/\,\,Hofbauer, J.\,/}\linebreak\pagebreak
\par {\em Wolff, R.} (1981),                             
Selfregulation of behavior in animal societies. Biological 
Cybernetics {\bf 40}, 1-25
\bibitem{Verhulst}{\em Verhulst, P. F.} (1845), Nuov. Mem. Acad. Roy. 
 Bruxelles {\bf 18}, 1
\bibitem{Volterra}{\em Volterra, V.} (1931), Le\c{c}ons sur la th\'{e}orie
 math\'{e}\-ma\-tique de la lutte pour la vie. Paris: Gauthier-Villars
\bibitem{Weid1}{\em Weidlich, W.\,/\,\,Haag, G.} (1983), Concepts and Models of a
 Quantitative Sociology. The Dynamics of Interacting Populations.
 Berlin: Springer
\bibitem{Mig}{\em Weidlich, W.\,/\,\,Haag, G.} (1988), Interregional Migration.
 Berlin: Springer
\bibitem{Wei}{\em Weidlich, W.} (1991), Physics and social science---The approach
 of synergetics. Physics Reports {\bf 204}, 
 1-163
\bibitem{Zipf}{\em Zipf, G. K.} (1946), The P1P2/D hypothesis on the intercity movement
 of persons. American Sociological Review {\bf 11}, 677-686
\end{document}